\documentclass[aps,prx, amsmath,amssymb,twocolumn,superscriptaddress]{revtex4-2}

\usepackage{graphicx}
\usepackage{bm}
\usepackage{hyperref}
\usepackage{dsfont}
\usepackage{color}
\def\Tr{\mathop{\mathrm{Tr}}}
\usepackage{ulem}
\usepackage{xcolor}

\usepackage{xspace}

\renewcommand{\i}{\boldsymbol{i}}
\renewcommand{\j}{\boldsymbol{j}}

\newcommand{\ve}[1]{\boldsymbol{#1}}

\begin{document}

\title{Hubbard and  Heisenberg models on hyperbolic lattices: Metal-insulator transitions, global antiferromagnetism, and  enhanced  boundary  fluctuations }
\author{Anika  G\"otz}
\affiliation{\mbox{Institut f\"ur Theoretische Physik und Astrophysik,
		Universit\"at W\"urzburg, 97074 W\"urzburg, Germany}}
\author{Gabriel Rein}
\affiliation{\mbox{Institut f\"ur Theoretische Physik und Astrophysik,
    Universit\"at W\"urzburg, 97074 W\"urzburg, Germany}}
\author{Jo\~{a}o Carvalho Inácio}
\affiliation{\mbox{Institut f\"ur Theoretische Physik und Astrophysik,
		Universit\"at W\"urzburg, 97074 W\"urzburg, Germany}}
\author{Fakher F. Assaad}
\affiliation{\mbox{Institut f\"ur Theoretische Physik und Astrophysik,
    Universit\"at W\"urzburg, 97074 W\"urzburg, Germany}}
\affiliation{W\"urzburg-Dresden Cluster of Excellence ct.qmat, Am Hubland, 97074 W\"urzburg, Germany}

\date{\today}

\begin{abstract}
We study the Hubbard and Heisenberg models on hyperbolic lattices with open  boundary  conditions by means of  mean-field approximations, spin-wave theory, and quantum Monte Carlo (QMC) simulations.   For the Hubbard model we  use  the auxiliary-field  approach 
and  for   Heisenberg systems  the stochastic series expansion algorithm
and   concentrate on  bipartite lattices where the QMC simulations  are free  of  the 
negative  sign problem.  The hyperbolic lattices have an extensive number of sites on the boundary, such that one has to distinguish between bulk  and total density of states.
The  considered  lattices  are  characterized by  a  Dirac-like total density of states,  Schläfli indices $\{p,q\}=\{10,3\}$ and  $\{8,3\}$, as  well as   by flat  bands, $\{8,8\}$. The  Dirac total density of  states  cuts off  the   logarithmic divergence  of  the  staggered spin susceptibility  and  allows  for  a finite $U$  metal-to-insulator  transition. This transition has the same  mean-field exponents  as  for  the  Gross-Neveu transition in  Euclidean space. We argue that this
transition is induced by the open-boundary conditions and that it will be absent in the periodic case.  In the presence  of flat  bands we  observe  the  onset  of  magnetic ordering at  any finite $U$.  This conclusion holds
even though the bulk density of states is constant at the Fermi energy for the three considered lattices.
The  magnetic  state  at  intermediate 
coupling  can  be  described  as a  global antiferromagnet.  It  breaks the $C_p$ rotational 
and time-reversal symmetries but  remains invariant  under combined $C_p \mathcal{T}$  transformations.  The state is  characterized by  macroscopic  ferromagnetic moments, 
that  globally cancel.     We  observe  that  fluctuations on the  boundary of  the 
system  are  greatly  enhanced:  While  spin-wave calculations predict  the 
breakdown of  antiferromagnetism  on  the boundary  but  not  in the bulk,  QMC simulations  show a marked reduction of  the  staggered moment on the edge of  the system.
\end{abstract}

\maketitle
\section{Introduction}
The effect of  dimensionality and  point group  symmetries on correlation induced phenomena is  an established research domain in flat 
Euclidean  space.  The aim of  this paper is to  investigate 
the effect curvature  has  on the physics  of standard models of  correlated electrons and spins. 
The role of the curvature is widely studied in models of statistical mechanics, like the Ising model \cite{Wu96,Wu00,Shima06,Shima06a,Ueda07,Sakaniwa09,Daniska16,Breuckmann20}. A constant negative curvature can for example lead to the presence of symmetry-broken phases and phase transitions, not present in the flat Euclidean plane \cite{Wu96,Benjamini99,Wu00,Benjamini01,Baek09a} or change the critical properties of phase transitions \cite{Callan90,Shima06,Ueda07}.
Also in condensed matter physics the popularity of hyperbolic lattices is increasing. The topics of interest are both single-particle physics, like the electronic band structure  \cite{Kollar20,Yu20,Boettcher20,Maciejko21,Ikeda21,Maciejko22,Boettcher22,Stegmaier22,Zhang22,Liu22,Bzdusek22,Urwyler22,Chen23a,Gluscevich23,Mosseri23}---also under the influence of a magnetic field and  topological phenomena---and correlation effects \cite{Zhu21,Bienias22,Gluscevich23}. The motivation can partially be attributed to recent experimental realizations of hyperbolic lattices in circuit quantum electrodynamics and electric circuits \cite{Kollar19,Lenggenhager22}.

In this paper we study the Hubbard and  Heisenberg models on three exemplary regular tessellations of the hyperbolic plane, that exhibit different electronic properties in the noninteracting limit. 
We use a Hartree-Fock approximation, a spin-wave  analysis,  auxiliary-field  and  stochastic  series  expansion quantum Monte Carlo (QMC)  methods to investigate the models as a function of the interaction strength and system size. 
Since all considered lattices are bipartite, we can perform QMC simulations without a negative sign problem  for both  the Hubbard  and  Heisenberg models.  
There is a plethora of exotic phases of matter, such as spin liquids, that have been realized on two-dimensional bipartite lattices \cite{Balents02,Assaad16,Hohenadler18,Xu18}.  This provides a further motivation to  study  correlation effects on bipartite hyperbolic lattices.
We will concentrate on systems with open boundary conditions. This is important since the number of boundary sites is extensive, such that bulk and total density of states (DOS) differ. The mean-field approximation of the Hubbard model on the $\{10,3\}$ lattice is already studied in 
Ref.~\cite{Gluscevich23}, and we complement these results with both spin-wave calculations and first-time fermionic QMC studies on the hyperbolic lattice.

The remainder of the paper is organized as follows. In Sec.~\ref{sec:model} we define the Hubbard and  Heisenberg  models and the hyperbolic lattices  on which we 
study them.  In Sec.~\ref{sec:method} we introduce the mean-field approximation, spin-wave analysis and the QMC algorithms that will be used in this work.  In Sec.~\ref{sec:results} we present our numerical results before we conclude in Sec.~\ref{sec:conclusion}. Additional numerical data as well as details and derivations of the methods used in this paper can be found in the Appendix.

\section{Hyperbolic Lattices, Hubbard and Heisenberg Models}\label{sec:model}
\subsection{Hyperbolic lattices}
\begin{figure}[tb]
 \includegraphics[width=1\linewidth]{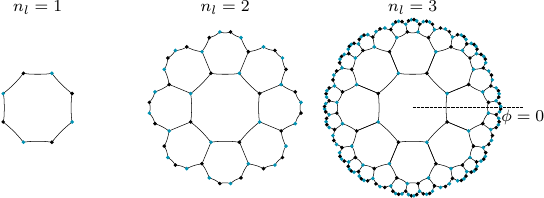}
  \caption{\label{fig:lattices}$\{8,3\}$ lattice in the Poincaré disk model with increasing number of layers $n_l$. Reference angle marked for latter use.   }
\end{figure}
In this work, we consider two-dimensional graphs that can be categorized by their Schläfli index $\{p,q\}$. The lattice sites are located at the corners of regular $p$-gons and every lattice site has coordination number $q$ \cite{Coexter73}. Hyperbolic lattices with a constant negative curvature fulfill the condition $(p-2)(q-2)>4$. In contrast, lattices with $(p-2)(q-2)=4$ lie within the flat Euclidean plane and  lattices with $(p-2)(q-2)<4$  on a sphere, that has a constant positive curvature. For the Euclidean plane exactly three solutions exist---the square lattice $\{4,4\}$, the honeycomb lattice $\{6,3\}$ and the triangular lattice $\{3,6\}$, and for the sphere five---the five Platonic solids, whereas for the hyperbolic case infinitely many pairs $\{p,q\}$ exist that fulfill the condition \cite{Edmonds82}. 

We concentrate on finite-size hyperbolic lattices with open boundary conditions. The lattice size can be increased by starting with a single $p$-gon and successively adding shells or layers of $p$-gons. The number of lattice sites grows exponentially with the number of layers, strongly limiting the accessible lattice sizes for our calculations. In Fig.~\ref{fig:lattices} the $\{8,3\}$ lattice is shown with up to $n_l=3$ layers. To properly depict the lattices, the Poincaré disk model is used. All  lattice sites are projected into the open unit disk and  all edges, connecting two nearest-neighboring sites, are geodesic lines. Those geodesics are circular arcs in the Poincaré disk model, that would hit the unit circle perpendicularly and they all have the same length within the hyperbolic metric
\footnote{The metric in the hyperbolic plane with a constant negative curvature of $-1$ is given by $ds^2=4\frac{dx^2+dy^2}{(1-x^2-y^2)^2}$. Here, $x$ and $y$ are the coordinates within the open unit circle in the Poincaré disk model.}
\cite{Cannon97,Anderson05}.

Drawing conclusions from the results on finite lattices about the properties in the thermodynamic limit is somewhat more complicated than in the Euclidean plane. For open boundary conditions the ratio of lattice sites at the edge to the total number of lattice sites converges to a finite value, so the behavior of the edge sites is not negligible even for large system sizes \cite{Shima06,Sakaniwa09,Urwyler22,Mosseri23}. If the bulk properties are of interest, then the lattice can be successively increased and the calculation can only be done for the inner bulk part, neglecting the outer shells \cite{Shima06,Shima06a}. Another possibility is to use periodic boundary  conditions. Some classes of hyperbolic tilings can be placed on compact surfaces with higher genus $g\geq2$ with periodic boundary conditions \cite{Sausset07,Breuckmann20,Zhu21,Maciejko22,Boettcher22}. In this work we use open boundary conditions and also take into account the boundary contributions.
\begin{table}[tb]
\caption{\label{tab:coord}Number of lattice sites for considered lattice geometries as a function of the number of layers $n_l$.}
\begin{center}
\begin{tabular*} {\linewidth}{@{\extracolsep{\fill}}  r  c  c  c}
	\hline \hline
	&	$\{10,3\}$ & $\{8,3\}$ & $\{8,8\}$ \\
			\hline
$n_l=1$ & 10 & 8 & 8 \\
$2$ & 80  & 48  & 56 \\
$3$ & 490  & 200 & 392 \\
$4$ & 2880  & 768 & 2744 \\
$5$ & 16810 & 2888 &  \\
$6$ &  & 10800 & \\
\hline \hline
	\end{tabular*}
	\end{center}
\end{table} 

Figure~\ref{fig:lattices} also shows, that the $\{8,3\}$ lattice is bipartite with sublattice $A$ and $B$, like all lattices in this work. In general for even $p$ the resulting lattices are bipartite \cite{Mosseri23}.
In Table~\ref{tab:coord} we provide a list of the considered lattice sizes and their respective number of lattice sites $N_s$. 

\subsection{Hubbard and Heisenberg models}
The Hamiltonian of the Hubbard model is given by \cite{Hubbard63,Gutzwiller63,Kanamori63}
\begin{eqnarray}
\hat{H} &=& \hat{H}_t + \hat{H}_U \,, \nonumber \\
 \hat{H}_t &=& -t \sum_{\langle\i,\j\rangle}  \sum_{\sigma=1}^{N} \left( \hat{c}_{\i,\sigma}^{\dagger} \hat{c}_{\j,\sigma}^{\phantom{\dagger}} + \mathrm{H.c.} \right) \,, \nonumber \\
\hat{H}_U &=&  \frac{U}{N} \sum_{\i} \left(\sum_{\sigma=1}^{N} \hat{n}_{\i,\sigma} -\frac{N}{2} \right)^2 \,. 
\end{eqnarray}
 The first term describes the hopping of fermions  
 on bonds $b=\langle \i,\j \rangle$ connecting nearest-neighboring sites $\i$ and $\j$ with hopping strength $t$.
The operator $\hat{c}_{\i,\sigma}^{\dagger}$ creates a fermion in a Wannier state centered at
site $\i$ and with $z$ component of spin equal to $\sigma$. We will consider $N=2$ (spin $S=\frac{1}{2}$) flavored fermions.
The fermions interact via a repulsive on-site Hubbard potential with strength $U$, as described in the second term.

The model is symmetric under global SU(2) rotations with the spin operators as the generators of the group
\begin{equation}
\hat{S}^{\alpha} = \frac{1}{2}\sum_{\ve{i},\sigma,\sigma'} \hat{c}^{\dagger}_{\bm{i},\sigma} \sigma^{\alpha}_{\sigma,\sigma'} \hat{c}^{\phantom{\dagger}}_{\bm{i},\sigma'} \, .
\end{equation} 
Here $\sigma^{\alpha=x,y,z}$ are the three Pauli matrices.
Furthermore,  the model  is invariant under  particle-hole transformations $\hat{P}^{-1} \hat{c}^{\dagger}_{\bm{i},\sigma} \hat{P} = (-1)^{\bm{i}} \hat{c}_{\bm{i},\sigma}^{\phantom{\dagger}}$  on a bipartite lattice at half-filling.
The prefactor $(-1)^{\bm{i}}$ is a multiplicative phase factor, that takes the value $+1$ on sublattice $A$ and $-1$ on sublattice $B$.

The lattice geometry can significantly influence the electronic band structure and therefore the behavior of the electrons, when exposed to interactions. This effect can be observed for the 
available regular bipartite graphs in the Euclidean plane, the square lattice $\{4,4\}$ and the honeycomb lattice $\{6,3\}$. 
On both lattices the repulsive Hubbard model at half-filling  undergoes a transition from a (semi-)metal to an antiferromagnetic (AFM) insulator in the limit of strong on-site interactions.
The Fermi surface of the half-filled square lattice exhibits van Hove singularities leading to a logarithmically divergent DOS at the Fermi surface. This results in an essential singularity of the staggered magnetization in the Hartree-Fock approximation, meaning that for any finite interaction strength $U$ the AFM state is expected to have a lower energy than the paramagnetic solution \cite{Hirsch85,Fradkin13}. Numerical studies support the presence of long-range AFM order for any finite interaction $U$ \cite{Hirsch85,Hirsch89,White89,LeBlanc15}.

On the honeycomb lattice the band structure of the noninteracting case exhibits two Dirac cones and the Fermi surface is concentrated on two points with a vanishing DOS. A finite critical on-site repulsion $U_c$ is needed for the formation of long-range AFM ordering \cite{Sorella92}. Numerical studies locate a continuous and direct semimetal to insulator transition at $U_c/t \approx 3.8$ \cite{Sorella12,Assaad13,Otsuka16}. This transition belongs to the Gross-Neveu universality class \cite{Herbut06,Assaad13,Otsuka16}.

For the hyperbolic case we know, that the unique ground state of the half-filled repulsive Hubbard model is a $S_{\mathrm{tot}}=0$ state for any bipartite lattice by a theorem of Lieb \cite{Lieb89}.
$S_{\mathrm{tot}}$ is the total spin and the theorem was proven for any bipartite collection of sites, meaning no explicit periodicity or dimensionality is assumed in the proof of the theorem. Furthermore, in the strong-coupling limit the model maps onto the Heisenberg model with an AFM coupling \cite{Anderson59}, since the mapping is independent of the lattice.   The  specific  form of  the model  reads: 
\begin{equation}
	\hat{H} = J \sum_{\langle \bm{i},\bm{j} \rangle } \hat{\bm{S}}_{\bm{i}} \cdot \hat{\bm{S}}_{\bm{j}} \,,
\end{equation} 
 with $J= \frac{4t^2}{U}$  and 
$ \hat{\ve{S}}_{\ve{i}} = \frac{1}{2}\sum_{\sigma,\sigma'} \hat{c}^{\dagger}_{\bm{i},\sigma} \ve{\sigma}_{\sigma,\sigma'} \hat{c}^{\phantom{\dagger}}_{\bm{i},\sigma'}$ .

\section{Methods}\label{sec:method}
\subsection{Mean-field approximation}\label{sec:mf}
As a first approach to study the Hubbard model on hyperbolic lattices we choose a mean-field approximation. We decouple the electron-electron interaction by coupling the $z$ component of the spin operator to the fields $\phi_{\bm{i}}$
\begin{eqnarray}\label{eq:Ham_MF}
\hat{H}_{\mathrm{MF}} &=& \hat{H}_t - U \sum_{\i} \phi_{\i} \left(\hat{n}_{\i,\uparrow} -\hat{n}_{\i,\downarrow} \right) + \frac{U}{2} \sum_{\i} \phi_{\i}^2 \,, \\
\phi_{\i} &=& \langle \hat{n}_{\i,\uparrow} -\hat{n}_{\i,\downarrow} \rangle_{\phi} = 2 \langle \hat{S}_{\bm{i}}^z \rangle_{\phi} = m^{z}_{\i} \, , \label{eq:mf_order}
\end{eqnarray}
where $\langle \dots \rangle_{\phi}=\frac{\Tr[\dots e^{-\beta\hat{H}_{\mathrm{MF}}}]}{\Tr[e^{-\beta\hat{H}_{\mathrm{MF}}}]}$ is the expectation value with respect to $\hat{H}_{\mathrm{MF}}$ and $\beta=T^{-1}$ the inverse temperature.
The fields $\phi_{\bm{i}}$ are equivalent to the local magnetic order parameter and due to the absence of translation symmetry in the considered lattices with open boundary conditions, we allow the fields to vary from site to site.  
In the following, Eq.~(\ref{eq:mf_order}) will be solved self-consistently. 
In Appendix \ref{sec:suscep} we  show  that,  due  to particle-hole 
symmetry, the  staggered  spin  susceptibility  diverges  logarithmically in the low-temperature limit,  the  prefactor  
being the DOS at the Fermi energy. 
Hence, we  foresee AFM ordering  and  define  the staggered order parameter
\begin{equation}
m^z = \frac{1}{N_s} \sum_{ \bm{i} } (-1)^{\bm{i}} m^z_{\bm{i}} \, .
\end{equation}

\subsection{Spin-wave approximation}
We continue with a spin-wave approximation. In the strong-coupling limit the Hubbard model maps onto the AFM Heisenberg model independently of the lattice geometry \cite{Anderson59}
\begin{equation}\label{eq:heisenberg}
\hat{H} = \sum_{\langle \bm{i},\bm{j} \rangle } \hat{\bm{S}}_{\bm{i}} \cdot \hat{\bm{S}}_{\bm{j}} = \frac{1}{2} \sum_{\bm{i},\bm{j}} T_{\bm{i},\bm{j}} \hat{\bm{S}}_{\bm{i}} \cdot \hat{\bm{S}}_{\bm{j}} \,.
\end{equation}
The matrix $T$ is the adjacency matrix of the lattice with $T_{\bm{i},\bm{j}}=1$, if sites $\bm{i}$ and $\bm{j}$ are nearest neighbors and $T_{\bm{i},\bm{j}}=0$ otherwise.
Following Ref.~\cite{Holstein40}, we use a Holstein-Primakoff transformation to map the spin operators $\hat{\bm{S}}_{\bm{i}}$ to boson operators $\hat{b}_{\bm{i}}^{\dagger}$ and expand in $\frac{1}{S}$. Since we start from an AFM, we choose on sublattice $A$ the representation 
\begin{equation}\label{eq:spin_a}
\hat{S}_{\bm{i}}^z = S - \hat{b}_{\bm{i}}^{\dagger}\hat{b}_{\bm{i}}^{\phantom{\dagger}} \, , \quad 
 \hat{S}_{\bm{i}}^{+} =  \hat{S}_{\bm{i}}^x+i\hat{S}_{\bm{i}}^y = \sqrt{2S} \hat{b}_{\bm{i}}^{\phantom{\dagger}}\,,
\end{equation}
and on sublattice $B$,
\begin{equation}\label{eq:spin_b}
\hat{S}_{\bm{i}}^z =  \hat{b}_{\bm{i}}^{\dagger}\hat{b}_{\bm{i}}^{\phantom{\dagger}} -S \,, \quad
 \hat{S}_{\bm{i}}^{+} =  \sqrt{2S} \hat{b}_{\bm{i}}^{\dagger} \, .
\end{equation}
The correction to the classically expected Néel state is given by the expectation value of the bosonic occupation number $\langle  \hat{b}_{\bm{i}}^{\dagger}\hat{b}_{\bm{i}}^{\phantom{\dagger}}  \rangle$. In Appendix~\ref{sec:app_spin} we sketch the calculation of the site-dependent correction.

\subsection{Quantum Monte Carlo}

To acquire deeper information about the electronic correlations on the various lattice geometries and support the mean-field  and  spin-wave data, we employ QMC studies  as our secondary approach. In the weak-coupling limit, we use a finite-temperature version of the auxiliary-field QMC (AFQMC) method to simulate the Hubbard model \cite{Blankenbecler81,White89,Assaad08_rev}. For the strong-coupling regime, the Hubbard model is mapped to an AFM Heisenberg model, and in this regime we use the stochastic series expansion (SSE) method with directed loop updates \cite{sandvik_1991, syljuasen_2002}.

\subsubsection{Auxiliary-field QMC}

 We use the Algorithms for  Lattice  Fermions (ALF) \cite{ALF22} implementation of
 the  finite-temperature  AFQMC  algorithm \cite{Blankenbecler81,White89,Assaad08_rev}.  It allows us to accurately simulate the model in various parameter regimes.  As for  the  Hubbard model  at  half-filling,   our  simulations do not  suffer  from the negative sign problem. 

Although the construction of momentum space, similar to Euclidean lattices, is possible \cite{Maciejko21,Maciejko22,Boettcher22}, it does not apply to our lattices with open boundary conditions due to the absence of translation invariance and we have no notion of momentum space. Yet, the QMC algorithm can be employed to calculate the uniform susceptibility
\begin{equation}\label{eq:uniform_suscep}
	\chi_O=\frac{\beta}{N_{s}}\sum_{\i,\j}^{N_{s}} \left(\big\langle \hat{O}_{\i} \hat{O}_{\j}\big\rangle -  \big\langle \hat{O}_{\i} \big\rangle \big\langle  \hat{O}_{\j}\big\rangle  \right)
\end{equation}
at nonzero Hubbard $U$ with $\hat{O}_{\i}$ being either density $\hat{n}_{\i}=\sum_{\sigma}\hat{n}_{\bm{i},\sigma}$ or spin operators $\hat{S}_{\i}^z$. To directly compare the QMC results with mean-field, we make use of the bipartite nature of the chosen lattices by constructing an AFM order parameter as
\begin{equation}
	m_{\rm{AFM}}^z = \sqrt{\frac{1}{N_{s}^2}\sum_{\i,\j}^{N_{s}}(-1)^{\i}(-1)^{\j}\big\langle \hat{S}_{\i}^z \hat{S}_{\j}^z\big\rangle}\,,
	\label{eq:order_QMC}
\end{equation} 
which is directly related to the spin-spin-correlation function. For all of the presented data, we used a code for the Hubbard model with SU(2)-decoupling and an imaginary time discretization
 of $\Delta\tau t=0.1$ \cite{ALF22}.

\subsubsection{Stochastic series expansion}

As a second QMC approach, we simulate the Heisenberg model on the hyperbolic lattices using the SSE method with directed loop updates \cite{sandvik_1991, syljuasen_2002}. Since the considered lattices are of bipartite nature, the method can be formulated without a sign problem. This approach allows us to access larger lattice sizes compared to the AFQMC method, since SSE simulations scale as \(\mathcal{O}(\beta N_s)\) as opposed to \(\mathcal{O}(\beta N_s^3)\) \cite{ALF22}. 

The SSE method is based on the Taylor expansion of the partition function \(Z\) in some complete basis \(\{\left| \alpha \right>\}\) (here we choose the \(S^z\) basis):
\begin{equation} 
	Z = \sum_{\alpha} \sum_{n=0}^{\infty} \sum_{S_n} \frac{(-\beta)^n}{n!} \left< \alpha \right| \prod_{i=1}^n H_{a_i, b_i} \left| \alpha \right> \,,
\end{equation}
where \(n\) is the current expansion order, \(S_n\) specifies the operator string \([(a_1, b_1), \hdots, (a_n, b_n)]\), and \(H_{a_i, b_i}\) is a bond term of the Hamiltonian which operates on bond \(b_i\). This action can be diagonal or off-diagonal depending on \(a_i\). The configuration space then consists of operator strings of varying size \(n\). To sample this space, we use two main types of updates. Diagonal updates are a local type of update, in which we insert/remove a Hamiltonian operator in \(S_n\). The directed loop update \cite{syljuasen_2002, syljuasen_2003} is a global type of update in which a loop is constructed and flipped in the vertex representation of \(S_n\), allowing us to change a large part of the configuration in one step. 

Both the uniform spin susceptibility [Eq.~\eqref{eq:uniform_suscep}] and the AFM order parameter [Eq.~\eqref{eq:order_QMC}] rely on the calculation of equal-time two-point spin correlations \(\big\langle \hat{S}^z_{\i} \hat{S}^z_{\j} \big\rangle\). As these operators are diagonal in the \(S^z\) basis, their evaluation within the SSE formalism is trivial.

\begin{figure*}[tb]
\includegraphics[width=1\linewidth]{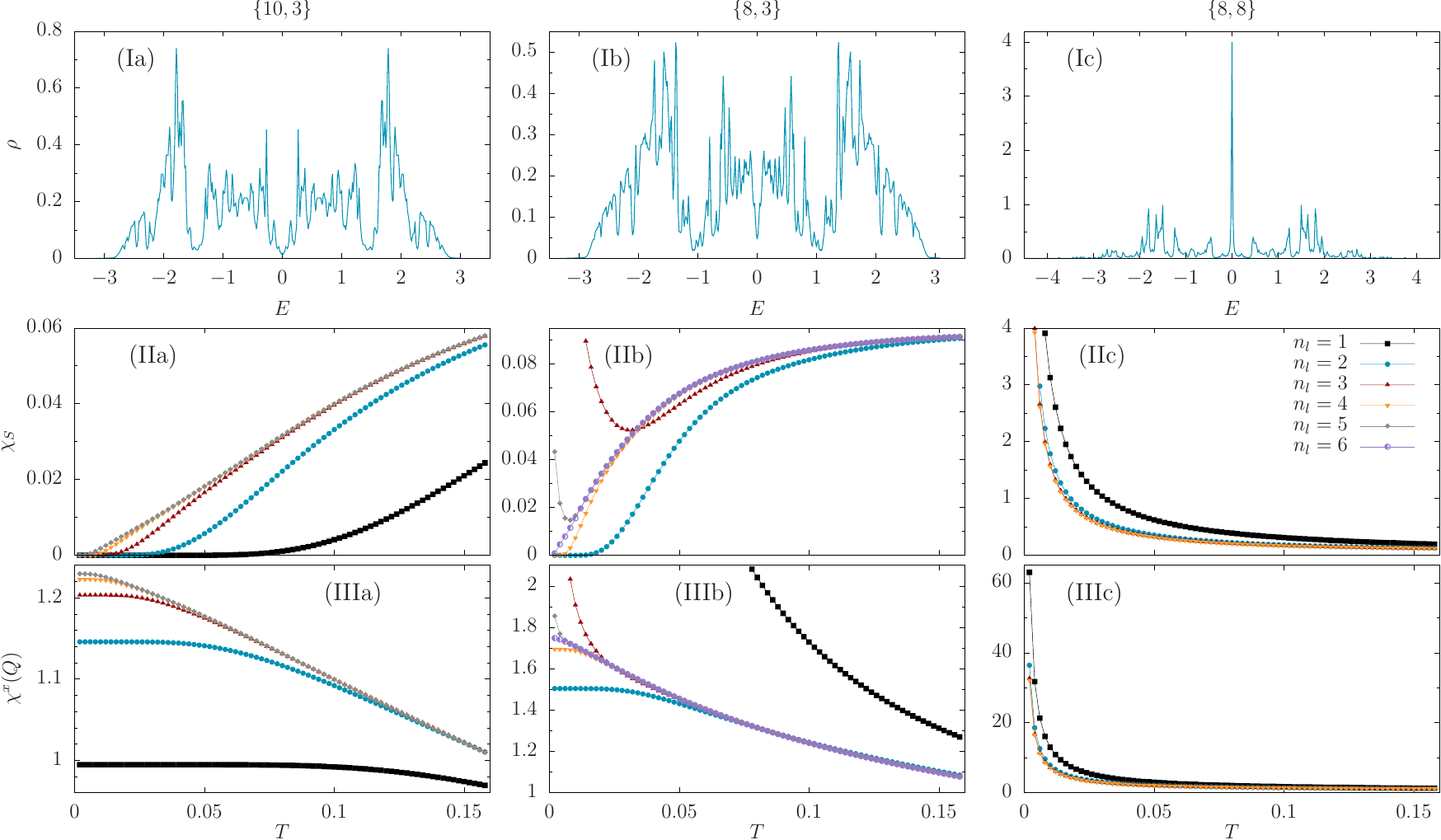}
  \caption{\label{fig:tight_bind}(Ia)--(Ic) Total DOS $\rho(E)$ as a function of the energy $E$ at $\delta=0.01$ for the largest considered lattice sizes ($n_l=5$ for the $\{10,3\}$ lattice, $n_l=6$ for $\{8,3\}$ and $n_l=4$ for $\{8,8\}$, see also Table~\ref{tab:coord}). (IIa)--(IIc) Uniform spin susceptibility $\chi_S$ and (IIIa)--(IIIc) staggered spin susceptibility $\chi^x(Q)$ as a function of the temperature $T$ and the number of layers $n_l$ in the noninteracting case $U=0$. }
\end{figure*}

\section{Results}\label{sec:results}
In the following we study the  SU(2) spin-symmetric Hubbard  and  Heisenberg models on three different hyperbolic lattice geometries at  the particle-hole symmetric point. As  shown in Appendix \ref{sec:suscep} particle-hole  symmetry  leads  to a  logarithmic 
divergence of  the  staggered spin  susceptibility, provided  that  the  DOS is  finite  at  
the  Fermi energy.   Hence,  the  first question that we will  address in Sec.~\ref{sec:u=0} is  the   DOS in the noninteracting limit. 
In Sec.~\ref{sec:mean_field} we present  and  discuss our  mean-field 
results.  In Sec.~\ref{sec:fluctutions},  we  then  systematically  take  into  account fluctuations   with a  spin-wave  approximation for  the Heisenberg model,   and     numerically exact  QMC  simulations  for  both  models.

\subsection{Non-interacting limit $U=0$}\label{sec:u=0}
First, we consider the total  DOS of the $\{10,3\}$ lattice, 
\begin{equation}
	\label{eq:dos}
\rho(E)=-\frac{1}{\pi N_s} \sum_n \mathrm{Im} G^{R}(n,E) \,.
\end{equation}
The retarded Green function is given by $G^{R}(n,E)=  \frac{1}{E-E_n+i\delta}$ with an infinitesimal $\delta$ and
  $E_n$ are the discrete energy eigenvalues of the tight-binding Hamiltonian $\hat{H}_t$.   Since  our  Hamiltonian enjoys  particle-hole symmetry, the  DOS  is   symmetric  around  $E=0$ and the  chemical potential  for   half-band   filling  is pinned to zero  for all  temperatures.
  The uniform spin susceptibility  $\chi_S$, defined in Eq.~(\ref{eq:uniform_suscep}), is  given  by 
  \begin{equation}\label{eq:spin_uniform}
		\chi_S = - (2S+1)\int  \,  dE \,  \rho(E) \frac{\partial f(E)}{\partial E}
  \end{equation}
  for  the free  case.   In Eq.~(\ref{eq:spin_uniform}),  $f$ denotes the Fermi function. 
  Owing to particle-hole symmetry, the staggered spin susceptibility,
  see Appendix \ref{sec:suscep}, reads:
\begin{equation}
  \label{eq:chi_xi_1}
  \chi^x(Q)  =  \int d E \frac{\rho(E)}{E} \tanh{\left(\frac{\beta}{2}E\right)}. 
 \end{equation}
Thereby, both $\chi^x(Q)$ and $\chi_S$   depend  upon the  total DOS.  If  $\rho(E) \propto |E| $, then
$\chi_S \propto T$  and $\chi^x(Q)$ converges to a constant.

The DOS,  uniform  and staggered  spin susceptibilities   are  plotted in Figs.~\ref{fig:tight_bind}(Ia)--\ref{fig:tight_bind}(IIIa) for the $\{10,3\}$ lattice.   The same quantities are considered in the (b) panels of Fig.~\ref{fig:tight_bind} for the $\{8,3\}$  lattice.  For these lattices  our results suggest a  Dirac-like total DOS, but finite-size effects   remain too
dominant for a definite conclusion. Size effects are smeared out by temperature. Let us first concentrate on the $\{8,3\}$  lattice.
At even number of layers, there are no zero-energy eigenstates in the spectrum and $\chi_S$ decreases 
exponentially in the low-temperature limit.  The  activation gap can be attributed   to the finite-size  gap.  
However, for an odd number of layers exactly two eigenstates at zero energy are present, thereby resulting in a $T^{-1}$ law of the uniform spin susceptibility in the low-temperature limit. But as the number of layers is increased the ratio of zero-energy eigenstates to the total number of eigenstates quickly converges to zero and the magnitude of the  $T^{-1}$  law vanishes in the large $n_l$ limit. 
For the  $\{8,3\}$   lattice,
   we can compare the data for the susceptibilities,  Figs.~\ref{fig:tight_bind}(IIb) and \ref{fig:tight_bind}(IIIb), for our
two biggest lattices,  $n_l = 5,6$ and observe convergence  down to $T/t \simeq 0.01$, a pretty cold temperature given the bandwidth of $6t$.  Extrapolation of the data  from $T/t > 0.01$  down to $T=0$ suggest a  finite staggered spin susceptibility and a vanishing uniform one.   Very same conclusions 
can  be  drawn for the  $\{10,3\}$  lattice.
We note that the slope of the DOS around $E=0$ and of the spin susceptibility is steeper for the $\{8,3\}$ lattice compared to the former lattice.
 In   Euclidean space,  the  slope is  given by  the inverse of the Fermi velocity.

The $\{8,8\}$ lattice, however, shows flat band features on the considered finite lattices. A large number of energy eigenstates are located at zero energy [Fig.~\ref{fig:tight_bind}(Ic)] and the uniform $\chi_S$ and staggered $\chi^x(Q)$ spin susceptibilities diverge in the zero temperature limit [Figs.~\ref{fig:tight_bind}(IIc) and \ref{fig:tight_bind}(IIIc)].

 \begin{figure}[tb]
  \includegraphics[width=1\linewidth]{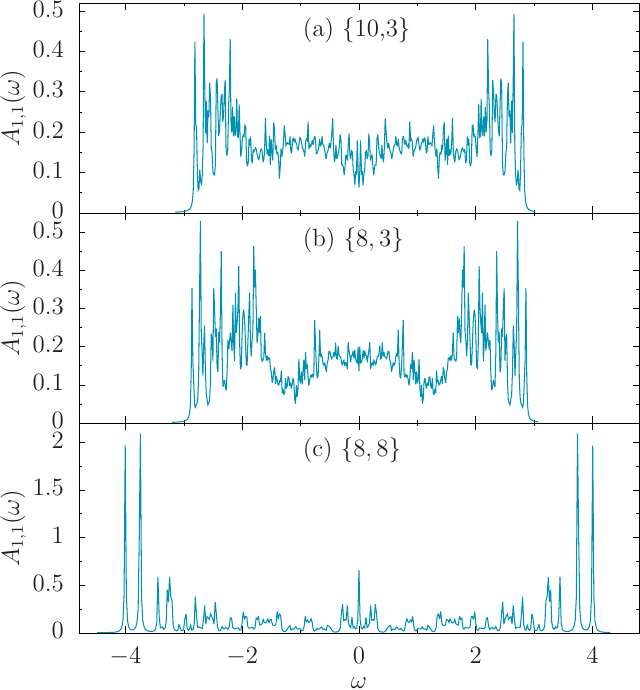}
   \caption{\label{fig:bulk_dos}Single-particle spectral function $A_{1,1}(\omega)$ at $\delta = 0.01$ and (a) $n_l=5$, (b) $n_l=6$, and (c) $n_l=4$. }
 \end{figure}

As mentioned  previously, the  boundary  of  hyperbolic lattices has  an  extensive number of  sites  such that  the bulk,   Ref.~\cite{Mosseri23},   and  total DOS, Eq.~\eqref{eq:dos}  can   differ.  
In Fig.~\ref{fig:bulk_dos}, we  consider  the  density of  states at a lattice site on the first layer, which we label as $n=1$.   The  bulk  density of  states, is then given by $A_{1,1}(\omega) = - \frac{1}{\pi} \text{Im}G^{R}(n=1,n'=1,\omega) $, where $G^{R}(n=1,n'=1,\omega)$ is  the  real-space retarded Green function   between points $n$ and $n'$  on the lattice.    As $n_l $ grows  this  quantity will capture the bulk DOS.  We plot this quantity in Fig.~\ref{fig:bulk_dos} for  the three lattices we consider  and on our largest system size.  As apparent, we can  reproduce  the results of
Ref.~\cite{Mosseri23} for the $\{10,3\}$ and $\{8,3\}$ lattices, and confirm that the bulk DOS takes a \textit{finite}   value at the Fermi energy for all considered lattices.

\subsection{Mean-field approximation}\label{sec:mean_field}

\subsubsection{Staggered magnetization and  metal-insulator transition}
\begin{figure}[htb]
\includegraphics[width=0.9\linewidth]{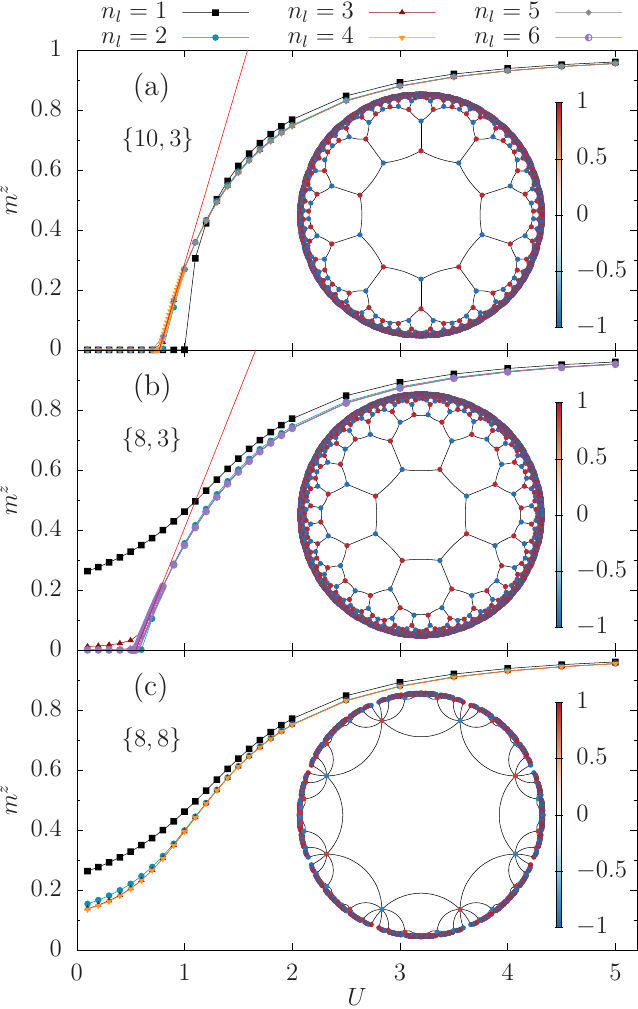}
  \caption{\label{fig:Mag_MF}(a)--(c) AFM order parameter $m^z$ as a function of the interaction strength $U$. The red line in panel  (a) is a linear fit to the $n_l=4$ data of the $\{10,3\}$ lattice close to the critical point $U_c\approx0.76t$ and in panel (b) to the $n_l=6$ data close to $U_c\approx0.55t$. Insets: local order parameter $m_{\bm{i}}^z$ at $U=5.0t$ for (a) $n_l=4$, (b) $n_l=5$, and (c) $n_l=4$. 
  } 
\end{figure}

\begin{figure}[tb]
	\includegraphics[width=1\linewidth]{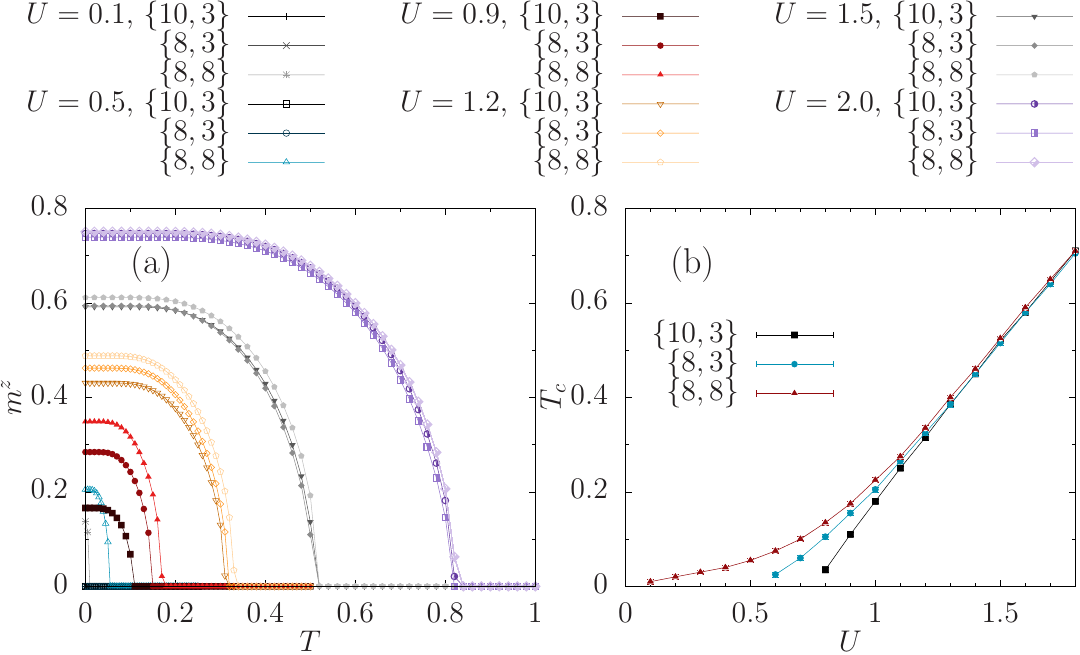}
	 \caption{\label{fig:mag_temp}(a) AFM order parameter $m^z$ as a function of the temperature $T$ for different interaction strengths as obtained from the mean-field approximation. (b) Critical temperature $T_c$ as a function of the Hubbard interaction $U$. For the $\{10,3\}$ and $\{8,8\}$ lattices we chose $n_l=3$ layers and for the $\{8,3\}$ lattice $n_l=4$ layers. 
	  }
   \end{figure}

In this section we study the three lattice systems with the mean-field approximation, introduced in Sec.~\ref{sec:mf}. 
Figure~\ref{fig:Mag_MF}  shows the order parameter as a function of the interaction strength $U$. In the strong-coupling limit a clear AFM state develops for all lattices with a vanishing total spin and a strong AFM moment $m^z$. The insets of Fig.~\ref{fig:Mag_MF} show the local magnetic moment $m_{\bm{i}}^z$, with a nearly spatially uniform modulus $|m_{\bm{i}}^z|$, deep in the AFM phase at $U=5t$.

The behavior of the three lattices in the weak-coupling limit differ strongly as a consequence of the total DOS in the noninteracting case.  In fact, in Appendix~\ref{sec:suscep} we show, that for a finite total DOS at the Fermi level and under the assumption of partial particle-hole symmetry an instability toward AFM can be expected even in the absence of translational symmetry.  Hence for the Dirac-like lattices we would not expect an AFM instability at small values of the interaction.   For the open hyperbolic lattices and since the bulk DOS is finite at the Fermi energy,  one  could question this conclusion.  In Appendix~\ref{sec:bulk}, we  present  a calculation to address the  instability of  the bulk.  Here, we  place  a  Hubbard-$U$  term  only on the
first layer  and  ask  the question if in the infinite-size limit  order will  occur at  infinitesimally
small couplings. The calculation shows  that  in this limit there is no instability. In fact, using  the  $C_p$ symmetry of the lattices, we can define $p$ unit cells each  with  $N_s/p$  orbitals  and  a $p$-valued $k$-vector.  In this notation, which we will also  use below,   the susceptibility associated  with introducing a staggered magnetic field on the first layer  reads (see also Appendix~\ref{sec:bulk}):
\begin{eqnarray}\label{eq:spin_staggered}
  \chi(Q)  &=  & 2 m^2  \sum_{k} \int d\omega d\omega'   \frac{f(-\omega')-f(\omega)}{\omega+\omega'}  \nonumber \\
   & & \times A_{1,1}(k,\omega)A_{1,1}(k,\omega') \,. 
\end{eqnarray}
Here  the spectral function $A(k,\omega)$ is a matrix  of dimension given by the number of
orbitals in the unit cell, and  the orbital labeled by 1  belongs to the first layer.  Equation~(\ref{eq:spin_staggered}) is generic. For a single band system, $ A_{1,1}(k,\omega)  = \delta(\omega - \epsilon(k)) $  
such that the generic log-divergence   is  reproduced.   However, in our case, when  the limit $n_l \rightarrow \infty $, $A_{1,1}(k,\omega)$  reflects the bulk DOS and takes a constant value  at the Fermi energy. Under those circumstances the integral does not diverge.   Hence  for 
the $\{10,3\}$  and  $\{8,3\}$ lattices,  we do not expect the onset of long ranged magnetic order at small values of  $U$.

For the Dirac-like $\{10,3\}$ lattice the order parameter $m^z$ indeed picks up a nonzero value only above a critical  interaction strength $U_c \approx0.76t$. Below $U_c$  the paramagnetic state is energetically favorable. 
Hartree-Fock mean-field theory for the flat honeycomb lattice predicts a linear increase of the order parameter above the critical interaction $m^z\propto (U-U_c)$, i.e., a critical exponent of $\beta=1$ \cite{Sorella92,Herbut09,Raczkowski20}. The order parameter of the $n_l=4$ lattice is well described by a linear fit close to the critical point [Fig.~\ref{fig:Mag_MF}(a)].

The  order parameter on the $\{8,3\}$ lattice  exhibits clear  odd-even effects in $n_l$. Let us consider the $n_l=1$ lattice,  which corresponds to the  $p$-site  ring with $p=8$. 
Since  the  rotational symmetry  of the ring  is  shared  by  the  $n_l> 1 $ lattices,  a  symmetry  analysis will  prove to be  very   useful.   The ground state of the  $p=8$  site  ring is    four fold degenerate: 
\begin{equation} 
	\label{eq:C8}
	  | \pm,  \sigma \rangle    =   \hat{c}^{\dagger}_{k = \pm \frac{\pi}{2}, \sigma}   | 0 \rangle \,.
\end{equation}  
In Eq.~(\ref{eq:C8}),  $ | 0  \rangle $ is  the  nondegenerate  ground  state  of   the  eight-site  ring   occupied with  six electrons,  and   
 $ \hat{c}^{\dagger}_{k,\sigma}   =   \frac{1}{\sqrt{p}}   \sum_{j = 1}^{p} e^{i k j } \hat{c}^{\dagger}_{j, \sigma}  $.  
Under    time-reversal  symmetry, $\mathcal{T}$,  and $C_8$ rotations  the  states  
transform  as 
\begin{equation}
       \mathcal{T}| \pm,  \sigma \rangle    =   \left( i \sigma^{y} \right)_{\sigma,\sigma'} | \mp,  \sigma ' \rangle
\end{equation}  
and  
\begin{equation}
	C_8| \pm,  \sigma \rangle    =   e^{\pm i \frac{\pi}{2} } | \pm,  \sigma  \rangle \,.
\end{equation}  
In the  presence  of  AFM  ordering,    time-reversal   and    $C_8$   symmetries  are  individually  broken,    but  the  combined  
$C_8 \mathcal{T}$  symmetry  is  conserved.    Terms  such as   
\begin{equation}
\Delta \sum_{\sigma}  \sigma   \left[ | +,\sigma\rangle  \langle - ,\sigma |    + | -,\sigma\rangle  \langle + ,\sigma |   \right]  
\end{equation} 
are  allowed  and  will  lift  the  ground-state degeneracy, yielding 
\begin{equation}
	| \Psi_0 \rangle  =    \frac{1}{2}\left( | +, \uparrow \rangle  + | -, \uparrow \rangle \right)   \otimes \left( | +, \downarrow \rangle  - | -, \downarrow \rangle \right) \,. 
\end{equation}
This  state  has  a  finite staggered magnetization $m^z$ for  any  value  of  $\Delta$,  and is  symmetric  under  combined   $C_8 \mathcal{ T}$  symmetries.  The  jump  in  the    magnetization  observed  at  $n_l = 1$   is  a  
consequence  of the  degeneracy  of   the  noninteracting  ground  state  
and the  concomitant    divergence  of  the uniform  spin susceptibility.   

While  the  symmetry  analysis  will apply to  arbitrary  values of  $n_l$, the    degeneracy 
  only occurs  for  odd  values of  $n_l$.  Since the degeneracy is  intensive,  the  jump in the  staggered    magnetization  will be suppressed as a function of  $n_l$.  For even   values of   $n_l$ the  finite-size  gap   leads  to  a  finite  value  of  $U\approx0.5t-0.6t$  beyond  which  magnetic  ordering will occur.  
For the largest system size in Fig.~\ref{fig:Mag_MF}(b) with $n_l=6$, the data can be described by a linear fit with $U_c\approx0.55t$ similar to the $\{10,3\}$ lattice.

In the  large  $n_l$ limit  we  conjecture  a Dirac-like  system.   The---in comparison to the $\{10,3\}$ lattice  lower---critical interaction strength, which lies within the range of $U_c\simeq0.5t-0.6t$, is consistent with the observation of Sec.~\ref{sec:u=0}, that the slope of the DOS around $E=0$ and of the uniform spin susceptibility of the $\{8,3\}$ lattice is steeper than the one of the $\{10,3\}$ lattice. Mean-field analysis for a Dirac system in flat space predicts a linear proportionality $U_c \propto v_{\mathrm{F}}$ \cite{Raczkowski20}.

In Fig.~\ref{fig:mag_temp}(a) we present the AFM order parameter as a function of temperature $T$,  and in Fig.~\ref{fig:mag_temp}(b) we present the  critical 
temperature  as  a function of the  interaction strength.  For  the  Dirac-like  systems   our  data is  consistent  with  $ m^{z}   \propto  (T - T_c)^{1/2} $,  in the vicinity  of  $T_c$  and   $T_c  \propto   (U -  U_c) $,  in the  vicinity of  $U_c$.   These   forms are  precisely  the  same as  obtained  for  Dirac  systems  on  Euclidean  lattices.
Hence  at  the mean-field  level, the transition in  hyperbolic space and  the  semimetal  to insulator transition   in flat  space are identical.

In contrast the flat-band-like $\{8,8\}$ lattice immediately supports AFM ordering for any finite $U$ due to the large DOS at the Fermi level in the noninteracting limit.    The singularity  at  $U=0$   for  all  considered  values of   $n_l$    can be  traced  back to the  divergence of  the uniform spin susceptibility and  accompanying   extensive   ground-state  degeneracy.   Although the  $T=0$  magnetization is   singular at  
$U=0$  the  N\'eel  transition  temperature  seems  to  follow  a  powerlaw 
[Fig.~\ref{fig:mag_temp}(b)] as  for  other  flat  band  systems  \cite{Bercx17}.  In  particular,  $ \lim_{T \rightarrow  0 } \lim_{U \rightarrow 0} m^z =  0 $,  but $ \lim_{U \rightarrow  0 } \lim_{T \rightarrow 0} m^z >  0 $.

\begin{figure}[tb]
 \includegraphics[width=1\linewidth]{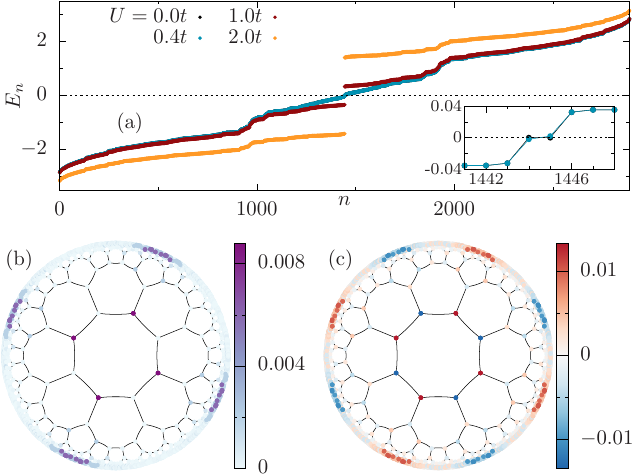}
  \caption{\label{fig:WF_global_AF}(a) Energy eigenvalues $E_n$ as obtained from the self-consistent solution of the mean-field Hamiltonian~(\ref{eq:Ham_MF}); inset: close-up of the Fermi level at $E_n=0$. (b) Absolute value of the wave function of  zero mode $|\psi_{n=1444}(\bm{i})|^2$ at $U=0.001t$ and $\sigma=\uparrow$ and (c) magnetization $m^z_{\bm{i}}$ at $U=0.4t$ for the $\{8,3\}$ lattice with $n_l=5$ layers.}
\end{figure}

\begin{figure}[tb]
	\includegraphics[width=1\linewidth]{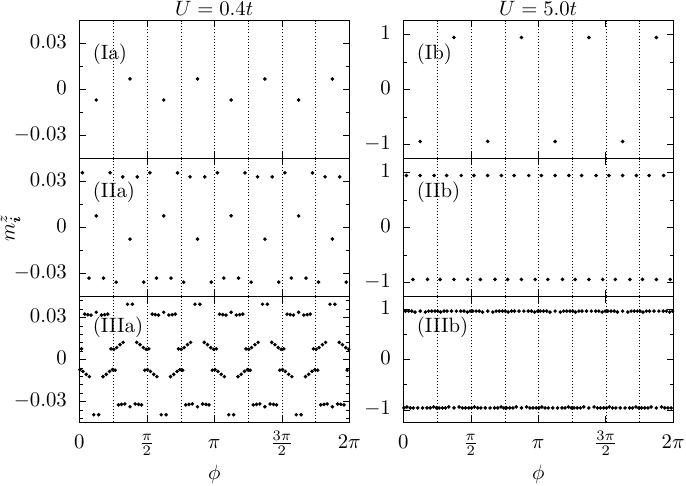}
	 \caption{\label{fig:angle_p8_q3}Local magnetization $m^z_{\bm{i}}$ as obtained from mean-field approximation as a function of the angle $\phi$ at two different interaction strengths (a) $U=0.4t$ and (b) $U=5t$ for the $\{8,3\}$ lattice with $n_l=3$. For clarity all sites of one layer are in a separate plot [(I) layer 1, (II) layer 2, and (III) layer 3]. 
	 }
   \end{figure}

\subsubsection{Global antiferromagnetism}
Using the $\{8,3\}$ lattice as an example, we now concentrate on the weak-coupling limit,
meaning that the interaction is (much) smaller than the electronic bandwidth in the 
noninteracting limit. Figure~\ref{fig:WF_global_AF}(a) shows the energy eigenvalues of the 
self-consistent solution of the mean-field Hamiltonian  (\ref{eq:Ham_MF}) for the $\{8,3\}$
lattice with $n_l=5$ layers and different interaction strengths $U$. In the noninteracting 
limit the spectrum is gapless with exactly two zero-energy eigenstates.
As for  the  $n_l=1$  case,  the  wave  function of  a  single zero 
mode, depicted in  Fig.~\ref{fig:WF_global_AF}(b), shows broken $C_8$ symmetry. 
Turning on a finite interaction opens a gap, [inset of Fig.~\ref{fig:WF_global_AF}(a)] and the AFM order parameter acquires a finite value.

As  for  the $n_l=1$ case, a small interaction strength of, e.g., $U=0.4t$ affects mainly the two eigenstates at the Fermi level [inset of Fig.~\ref{fig:WF_global_AF}(a)].  Figure~\ref{fig:WF_global_AF}(b) shows the absolute value of the wave function $|\psi_n(\bm{i})|^2$ of one zero mode in real space. To lift the degeneracy of the zero modes and have a uniquely defined pattern of the wave function, we choose a small interaction strength of $U=0.001t$ in   Fig.~\ref{fig:WF_global_AF}(b).
The wave function has support only on a small number of sites and it breaks the $C_{p=8}$ symmetry of the lattice down to $C_{p/2=4}$. The wave function of the second zero mode is similar, but rotated by $\pi/4$.

The spatial distribution of the order parameter $m^z_{\i}$ follows the pattern of these wave functions [Fig.~\ref{fig:WF_global_AF}(c)]. 
In particular,  at  finite  $U$  the   ground  state is unique,   breaks 
time-reversal  and    $C_8$   symmetries,  but it  satisfies: 
\begin{equation}
	 C_8\mathcal{T}  |\Psi_0 \rangle  =|\Psi_0 \rangle \,.
\end{equation}
For  the $n_l=1$ case,  this  just  results in  AFM,  since  the  $C_8$   transformation  amounts to  a unit  translation   along  the  chain.  
For   $n_l > 1 $,  new  magnetic structures  emerge,  which we   will  identify with  global  AFM in these 
 hyperbolic  lattices as   seen, e.g., in  strained graphene \cite{Roy14}. 

First,   the total   magnetic moment  vanishes.    However,  the  vanishing of  this  moment  involves cancellation of   extensive  ferromagnetic 
moments on    each   $p$-th sector of the lattice.
This is  depicted  in  Fig.~\ref{fig:angle_p8_q3}, where  the local magnetization $m_{\bm{i}}^z$ 
is plotted as a function of the angle $\phi$ in the different layers, both in the weak $U=0.4t$ 
and strong-coupling limit $U=5t$ for the $\{8,3\}$ lattice with $n_l=3$ layers. In the strong-coupling 
limit the magnetization is almost uniform on the whole lattice. At weak  coupling 
the  magnetization  varies strongly with the angle.  Summing  over  all  sites 
included  in  $\phi \in  [0,\pi/4[ $ produces  a  net  positive  result,  i.e.,  a 
ferromagnetic  moment  that cancels with  the contribution of  the sites included in
$\phi \in  [\pi/4, \pi/2[ $. 
The effect  is very prominent for the $\{8,3\}$ lattice with $n_l=5$
layers [Fig.~\ref{fig:WF_global_AF}(c)],
but we observed similar effects  in other lattice geometries 
as well, also in the absence of zero modes (see Appendix~\ref{sec:add_mean} 
for more examples). 

It is  tempting  to  identify  our   magnetic   system as an   altermagnet \cite{Smejkal22}.  A corner  stone  of an altermagnet is   compensated  magnetism  with spin-split  bands.    We can use   the  $C_p$  symmetry  to identify   a  unit  cell,  $i$,   with  an  orbital structure and 
corresponding   Brillouin zone. 
Using this notation,  the fermion creation operator  can  be  written as 
$\hat{c}^{\dagger}_{i,n,\sigma} $  where  $n$   denotes  orbital  $n$ in unit cell  $i$.  In this  
notation,   the   mean-field  Hamiltonian reduces  to: 
\begin{eqnarray}\label{eq:altermagnet}
  \hat{H}_{\mathrm{MF}}    =  & &    \sum_{i,i', n,n', \sigma}  \hat{c}^{\dagger}_{i,n,\sigma} T_{n,n'}(i-i')\hat{c}^{ \phantom\dagger}_{i',n',\sigma}    \nonumber  \\ 
      + & &  \sum_{i,n,n'\sigma}  \sigma (-1)^{i} 
  \hat{c}^{\dagger}_{i,n,\sigma} M_{n,n'} \hat{c}^{\phantom\dagger}_{i,n',\sigma}  \,.
\end{eqnarray}
In Eq.~(\ref{eq:altermagnet}), $M$ is  a  diagonal  matrix  that  encodes  the  site  magnetization in the unit  cell.  The factor  $(-1)^{i}$  accounts  for  the global  AFM. We  can now  
carry out a  Fourier  transformation   to  obtain: 
\begin{eqnarray}\label{eq:altermagnet_ii}
\hat{H}_{\mathrm{MF}} 	 =   \sum_{k \in \mathrm{MBZ}, \sigma}  & &  \left( \hat{\ve{c}}^{\dagger}_{k,\sigma},  \hat{\ve{c}}^{\dagger}_{k+Q,\sigma} \right)  \nonumber \\
& & \begin{pmatrix}
	 T(k) &  \sigma  M  \\
	 \sigma M^{\dagger} &  T(k+Q)
\end{pmatrix}
\begin{pmatrix}
	\hat{\ve{c}}^{\phantom\dagger}_{k,\sigma} \\
	\hat{\ve{c}}^{\phantom\dagger}_{k+Q,\sigma} 
\end{pmatrix} \,. 
\end{eqnarray} 
In Eq.~(\ref{eq:altermagnet_ii}),  $Q$   is  the one-dimensional  AFM wave vector,  MBZ  refers 
to the  magnetic  Brillouin zone, and $\hat{\ve{c}}^{\dagger}_{k,\sigma} $ is  a  spinor 
accounting  for  the  orbital  degrees of  freedom.   Using the  determinant  identity  based 
on the  Schur   complement,  one will readily  show  that the  energy spectrum  does not 
depend on the spin  index.   As such,  we  cannot  understand   the  observed  global  
AFM in terms of  an altermagnet.

\begin{figure}[tb]
 \includegraphics[width=1\linewidth]{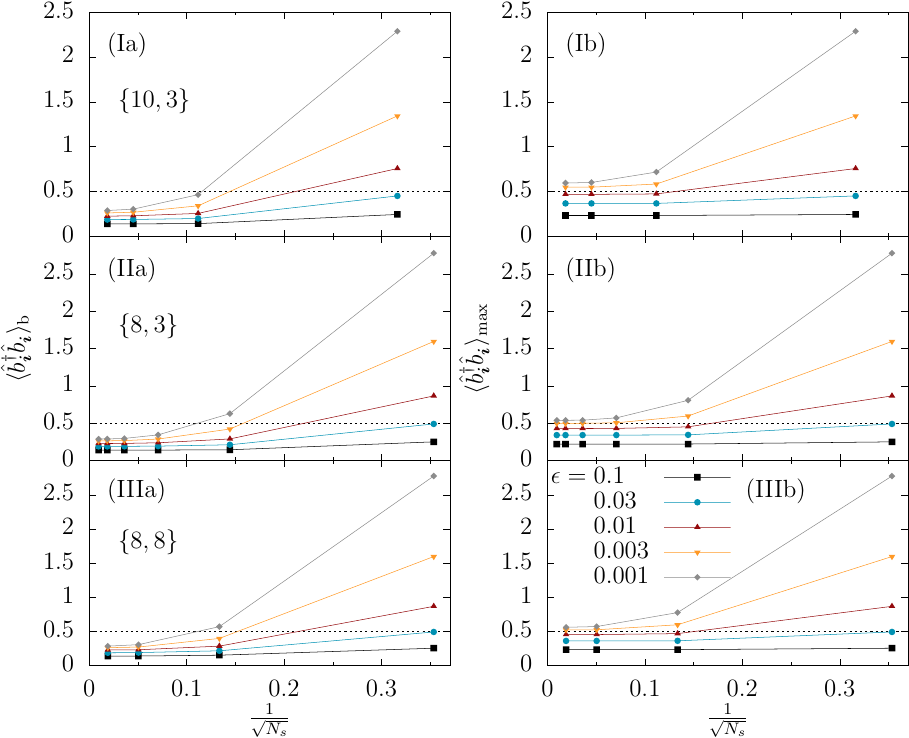}
  \caption{\label{fig:spin_wave}(a) Average correction $\langle \hat{b}_{\bm{i}}^{\dagger}\hat{b}_{\bm{i}}^{\phantom{\dagger}}\rangle_{\mathrm{b}}$ of the $n_l-1$ inner layers and (b) maximal correction   $\langle \hat{b}_{\bm{i}}^{\dagger}\hat{b}_{\bm{i}}^{\phantom{\dagger}}\rangle_{\mathrm{max}}$ as a function of the system size $\frac{1}{\sqrt{N_s}}$ at different strengths $\epsilon$ of the infinitesimal staggered field for the (I) $\{10,3\}$, (II) $\{8,3\}$, and (III) $\{8,8\}$ lattice. }
\end{figure}

\begin{figure}[tb]
	\includegraphics[width=0.8\linewidth]{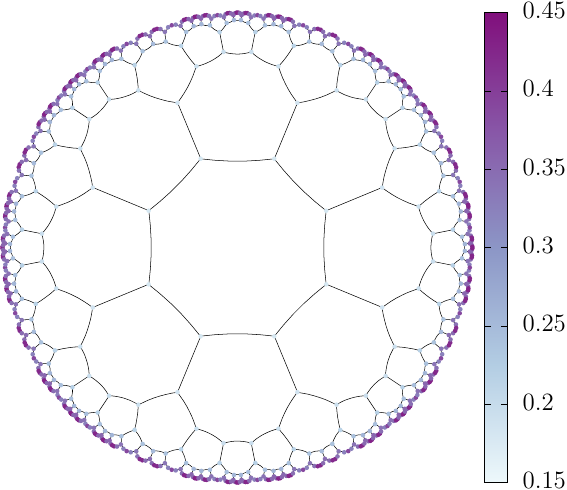}
	 \caption{\label{fig:spin_wave_local}Local correction $\langle \hat{b}_{\bm{i}}^{\dagger}\hat{b}_{\bm{i}}^{\phantom{\dagger}}\rangle$ to classical Néel state as obtained from spin-wave approximation for the $\{8,3\}$ lattice with $n_l=4$ at $\epsilon=0.01$.}
   \end{figure}

\subsection{Beyond  the mean-field approximation}\label{sec:fluctutions}

We  now  consider  approximate  as  well as  numerically  exact 
methods  that take into account  fluctuations.  We will  first   start with the spin-wave  theory of  the  Heisenberg model on 
hyperbolic  lattices,   and then  use  Monte  Carlo  methods for both 
the Hubbard and  Heisenberg models. 
\begin{figure*}[htb]
	\includegraphics[width=1\linewidth]{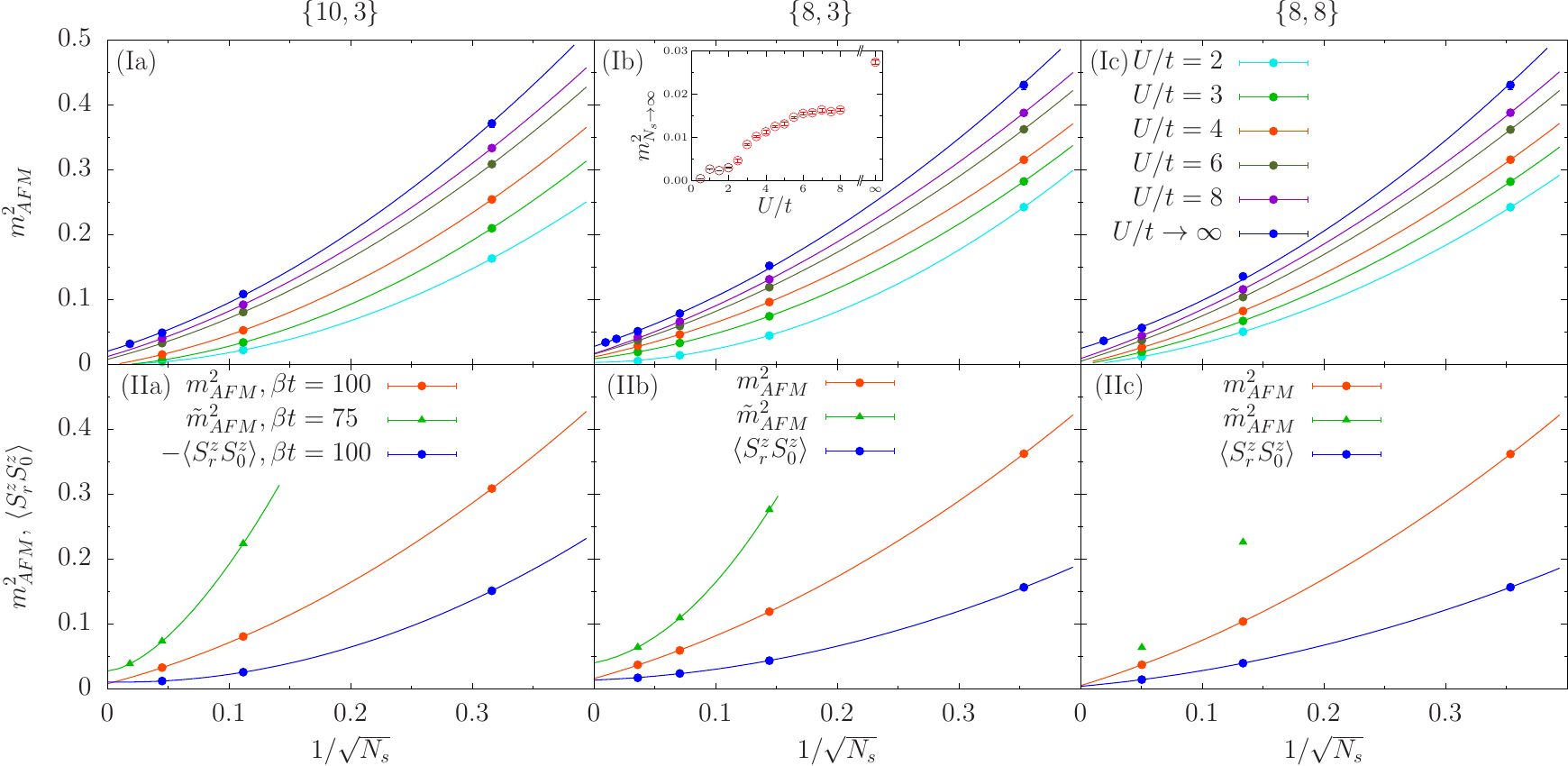}
	\caption{\label{fig:Mag_afm_QMC}(Ia)--(Ic) Finite-size scaling of the square of the AFM magnetization (order parameter), defined in Eq.~(\ref{eq:order_QMC}), at $\beta t=100$ for different values of $U/t$ on the $\{10,3\}$, $\{8,3\}$ and $\{8,8\}$ lattice respectively (from left to right). On the $x$ axis we plot the square root of the inverse number of lattice sites $N_s$. The data points are extrapolated to $N_s\to\infty$ by a  polynomial fit of second order and presented as a function of $U/t$ in the inset. We only present the extrapolated value for the $\{8,3\}$ lattice, since   we have  four data points  to carry out the fit.  We estimate that   the systematic error is of the order of  $0.005$  such  that  data points below this threshold    should be interpreted with care.
 The data points for $U\to\infty$ refer to the Heisenberg model at \(\beta J = 100\), calculated by SSE. (IIa)--(IIc) Comparison of the squared magnetization, the squared magnetization of the bulk $\tilde{m}_{\textrm{AFM}}^2$ and the spin-spin correlation $\langle S_{\bm{r}}^zS_0^z\rangle$ of two sites with the largest distance to each other on the respective lattice, where $S_0^z$ is the site closest to $\phi=0$ (for the $\{10,3\}$ lattice this is negative since opposing sites lie on different sublattices). Here, we use $U/t=6$.	}
\end{figure*}

\subsubsection{Spin waves}
The motivation to  carry out this  calculation stems  from the fact
that  the majority of  sites on the boundary of the hyperbolic
lattice  have a coordination number of  two akin to  a one-dimensional 
chain.  We hence  foresee that  fluctuations on the edge will be 
greatly  enhanced.  
To carry out  the calculation we  follow Refs.~\cite{Walker80,Sanyal12},   where spin-wave  calculations  were formulated  for  spin glasses.   
The method is  summarized in Appendix \ref{sec:app_spin}. To avoid   zero 
modes,  we  include a small staggered  field  of magnitude $\epsilon$ and  take the  limit  $\epsilon \rightarrow 0 $. 

In Fig.~\ref{fig:spin_wave} we show the correction to the classical Néel state in the $\frac{1}{S}$-expansion as obtained from Eq.~(\ref{eq:1/S}). Due to the absence of translation symmetry in the 
lattices, the corrections are site-dependent. The left panels show the average correction within the bulk $\langle \hat{b}_{\bm{i}}^{\dagger}\hat{b}_{\bm{i}}^{\phantom{\dagger}}\rangle_{\mathrm{b}}$, where we defined the bulk as the $n_l-1$ inner layers, and the right panels show the maximal correction $\langle \hat{b}_{\bm{i}}^{\dagger}\hat{b}_{\bm{i}}^{\phantom{\dagger}}\rangle_{\mathrm{max}}$, which can always be found on the sites closest to the angle $\phi=0$ in the outermost layer and all sites, that can be obtained by performing symmetry-allowed rotations. 

As described in Appendix~\ref{sec:app_spin} the limit of $\epsilon\rightarrow0$ and the thermodynamic limit have to be taken carefully. Taking the limit $\epsilon\rightarrow0$ on finite lattices leads to divergencies in the correction (Fig.~\ref{fig:spin_wave}) and only in the thermodynamic limit $\frac{1}{\sqrt{N_s}}\rightarrow0$ the correct results are recovered.

The average corrections in the bulk converge to values smaller than $\frac{1}{2}$ on the largest available lattice sizes for all three lattice geometries [Fig.~\ref{fig:spin_wave}(a)], meaning that the Néel state in the bulk is weakened, but still present in the case of spin $1/2$ degrees of  freedom. The maximal correction  $\langle \hat{b}_{\bm{i}}^{\dagger}\hat{b}_{\bm{i}}^{\phantom{\dagger}}\rangle_{\mathrm{max}}$, however,  is above $\frac{1}{2}$, such that it is large enough to destroy the magnetic moment on the corresponding sites.

In Fig.~\ref{fig:spin_wave_local} the local correction of the $\{8,3\}$ lattice with $n_l=4$ layers at $\epsilon=0.01$ can be seen. The correction $\langle \hat{b}_{\bm{i}}^{\dagger}\hat{b}_{\bm{i}}^{\phantom{\dagger}}\rangle$ depends strongly on the coordination number of the corresponding lattice site. In the bulk, where every site has $q=3$ neighbors, the correction is small. In the edge layer most sites are connected to only two neighboring sites, forming short one-dimensional sections. Therefore the correction on a given site is larger the greater the distance to the nearest site with $q = 3$ neighbors. This result is consistent with the absence of long-range order in one dimension, as  predicted by spin-wave theory.
Similar qualitative observations for the other lattice geometries can be found  in Fig.~\ref{fig:spin_wave_more_latt} of Appendix~\ref{sec:add_spin}.

\begin{figure*}[htb]
	\includegraphics[width=\linewidth]{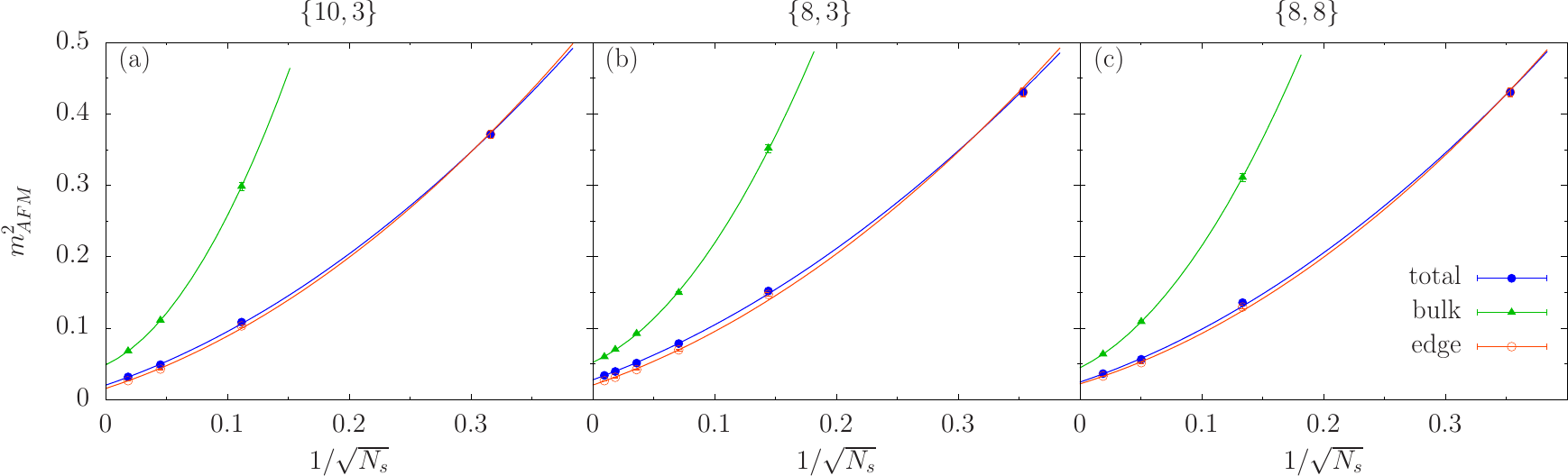}
	\caption{\label{fig:Mag_SSE} SSE calculated squared magnetization of the $\{10,3\}$, $\{8,3\}$, and $\{8,8\}$ lattice, respectively (from left to right), at \(\beta J = 100\), together with the bulk and edge contribution. The data points were fitted with a second order polynomial. }
\end{figure*}

\begin{figure}[b]
	\includegraphics[width=1\linewidth]{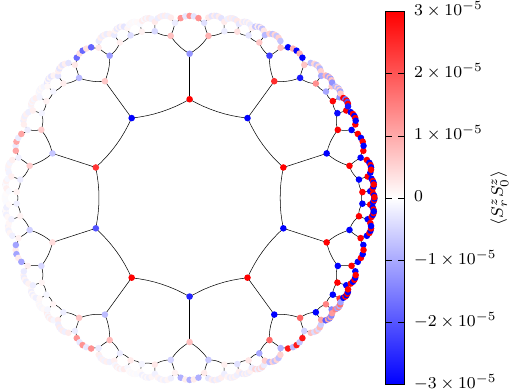}
	\caption{\label{fig:QMC_global_AF}Spin correlations $\langle S_{\bm{r}}^z S_0^z\rangle$ of the $\{10,3\}$ lattice for $n_l=3$ layers calculated by AFQMC at $U=1t$ and $\beta t=100$. As reference point $S_0^z$ we chose the site closest to $\phi=0$.}
\end{figure}
\begin{figure*}[htb]
	\includegraphics[width=1\linewidth]{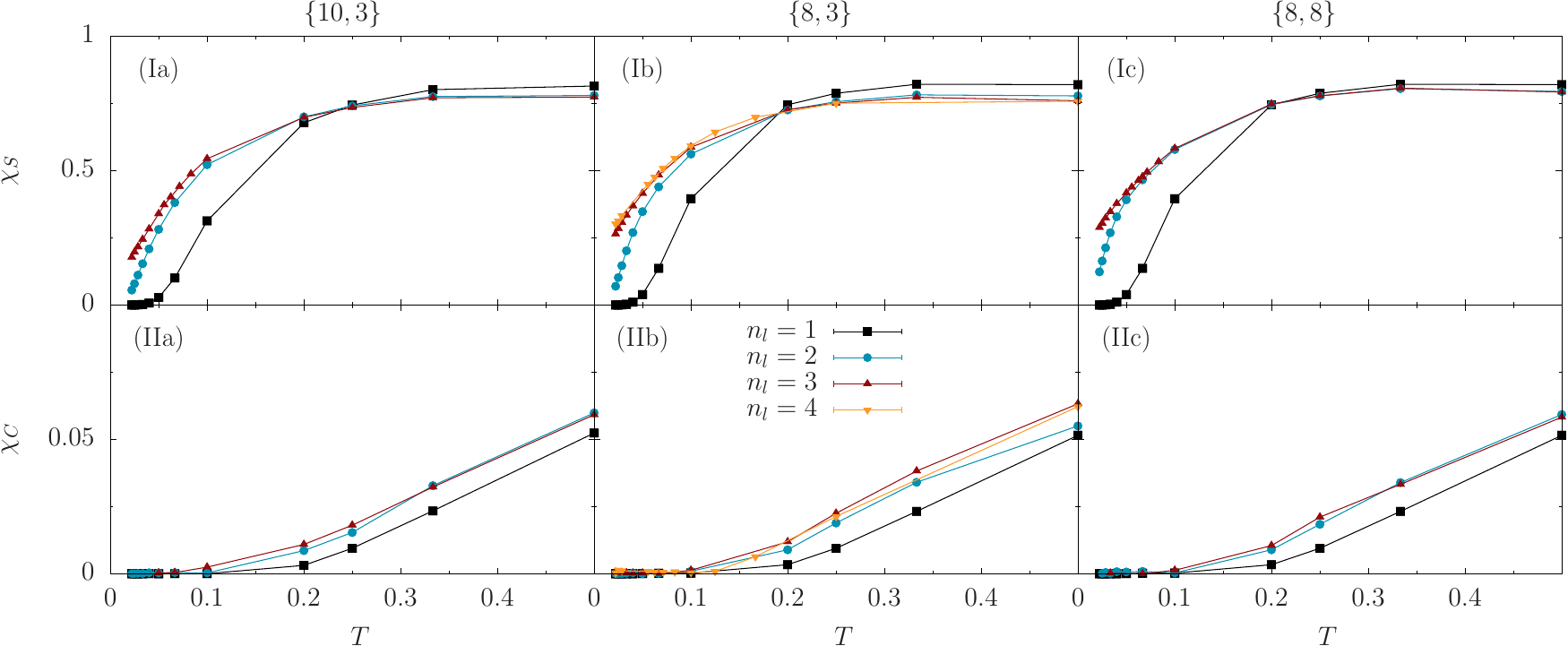}
	\caption{\label{fig:Susceptibility_QMC}QMC data for the (Ia)--(Ic) uniform spin susceptibility $\chi_S$ and (IIa)--(IIc) charge susceptibility $\chi_C$ at $U=5t$ on the $\{10,3\}$, $\{8,3\}$, and $\{8,8\}$ lattice, respectively (from left to right).}
\end{figure*}

\subsubsection{Quantum Monte Carlo}
The  aim of  this  section is to use  numerically  exact   approaches,  AFQMC  and  SSE,   to  check the validity of the  above statements.

Figures~\ref{fig:Mag_afm_QMC}(Ia)--\ref{fig:Mag_afm_QMC}(Ic)   plot the  staggered spin-spin correlations as defined in Eq.~(\ref{eq:order_QMC})  as  
a   function of $\frac{1}{\sqrt{N_{s}}}$. Unless  mentioned 
otherwise,  we consider  a finite, but  low  temperature  scale 
 $\beta t = 100$ and   fit the  data  to  the 
form $m^2_{N_s \rightarrow \infty} + a N_s^{-\frac{1}{2}} + b N_s^{-1}$. 
It is  notoriously  hard  to  extrapolate the  staggered  magnetization to  the 
thermodynamic  limit.  First, since  the number of  sites    grow  rapidly  with 
$n_l$   we  only  have a  limited set  of  data. Second,  we  measure  the  square  of  
the  order parameter  such  that when the  staggered  magnetization is  small (i.e., at  weak  coupling  or  in the proximity  of  a  quantum phase  transition)   very large  
lattices  are  required  to obtain reliable   results.    We note  that  a pinning 
field  approach \cite{Assaad13} that was  used  for  the  Hubbard model  on 
the honeycomb lattice  may  be  an  alternative  and  promising  method for  future 
investigations.  

For  the $\{10,3\}$ and $\{8,3\}$ lattices [Figs.~\ref{fig:Mag_afm_QMC}(Ia) and \ref{fig:Mag_afm_QMC}(Ib)]  our  data   at $\beta t = 100$ ($\beta  J = 100$) is 
representative  of  the  ground-state  properties  for  the considered  values of
$n_l$.   
In the strong-coupling limit, the results of both lattices are consistent with AFM long range ordering.
For  the  $\{8,3\}$  lattice  ordering  sets  in  at  $U_c \simeq 2t $.  
Upon inspection we see  that  the $\{10,3\}$   lattice  is still disordered  at $U/t = 3$. 
Hence, the   trends  in the  value  of  $U_c$,  $(U_c^{\{10,3\}} >  U_c^{\{8,3\}} $)
between the  two lattices  match  our  mean-field  analysis.  It  is  beyond   the scope 
of  this article to  determine the  critical exponent of the order  parameter: 
$m \simeq  (U -U_c)^{\beta} $.
In Figs.~\ref{fig:Mag_afm_QMC}(Ia) and \ref{fig:Mag_afm_QMC}(Ib),
we  equally  consider the  Heisenberg model.  For  this model,  the SSE  allows us  to
reach  much  larger  lattices   such  that   the extrapolation is more   robust. 

In Fig.~\ref{fig:Mag_afm_QMC}(Ic)  we  consider  the  flat  band  system $\{8,8\}$.  
With the  fermion QMC  we  can only  access  three  system  sizes for 
the extrapolation.  Hence,  we  do not  attempt to  determine  $m^2_{N_s \rightarrow 
\infty}$  for  this lattice.  Furthermore,  temperature effects   are  much  larger
for  this flat band system.   Nevertheless,  the data  is  consistent  with  ordering at
large  values  of  $U/t$  and   for  the  Heisenberg model.  

In Figs.~\ref{fig:Mag_afm_QMC}(IIa)--\ref{fig:Mag_afm_QMC}(IIc), we also present the AFM magnetization of the bulk $\tilde{m}^2_{\mathrm{AFM}}$ by only summing over sites of the $n_l-1$ inner layers. The bulk magnetization is larger than the total magnetization including the edge,  consistent with our results from spin-wave theory. This becomes more obvious when we consider the bulk and edge contribution to the magnetization calculated by SSE, as presented in Fig. \ref{fig:Mag_SSE}. This method allows a more accurate study of the magnetization in the strong-coupling limit as larger lattices can be simulated. While the edge magnetization is slightly lower than the total magnetization, the bulk contribution is larger by a factor of approximately $\sqrt{2}$ for all considered lattices. This specifies our findings from spin-wave theory that  fluctuations 
are  enhanced on the boundary. The SSE calculations suggest more precisely, that there is still long range order on the edge.  
Finally,  we  note  that  due  to the  subextensiveness  of  the bulk,  the magnetization  is  determined by  the  edge   contribution.  This is  supported  by  the SSE  data of Fig.~\ref{fig:Mag_SSE} where the  total  and  edge   data  scale  to the  same  value in the large $n_l$ limit.

The  global  AFM  is  a  weak-coupling effect,  such  that  we  use  the Hubbard model  and  AFQMC   for  investigations.  To compare the results from mean-field with QMC, we  present spin correlations plotted in the Poincaré disk of the $\{10,3\}$ lattice as shown in Fig.~\ref{fig:QMC_global_AF}. Again considering the weak-coupling regime, similar patterns of alternating positive and negative correlations can be observed in the third layer. This can be directly compared to the mean-field results of the order parameter for the $\{10,3\}$ lattice in Fig.~\ref{fig:global_afm_p10_q3}(a); see Appendix. For the presentation of the QMC results, we choose the $\{10,3\}$ lattice, since it offers the clearest visualization of the observed effect.   Note that  the  considered value of  $U/t$ is smaller  than  
the estimated  $U_c$.  Nevertheless,  the  observed  spin  fluctuations    support
the  notion  that in the  symmetry-broken phase  at large  $U/t$
the $C_{10}$  and  time-reversal   symmetry  are  broken,  but 
  $C_{10} \mathcal{T}$   
remains a   symmetry  of the state. 

To support the mean-field observation that the system is in an AFM Mott insulating  state for strong electronic correlations, we  present the spin and charge susceptibilities for $U=5t$ on the $\{10,3\}$, $\{8,3\}$, and $\{8,8\}$ lattices in Figs.~\ref{fig:Susceptibility_QMC}(Ia)--\ref{fig:Susceptibility_QMC}(Ic) and \ref{fig:Susceptibility_QMC}(IIa)--\ref{fig:Susceptibility_QMC}(IIc) respectively. As expected for an AFM state spin  waves lead  to a spin susceptibility $\chi_S$  that converges to a  finite value  as $T\to 0$. The data  shows  that there  is a  buildup of  low 
energy  states  as one increases  $n_l$. 
However, the  charge susceptibility $\chi_C$ decays exponentially, as  expected for a charged, gaped  system.    Figures~\ref{fig:Susceptibility_QMC}(IIa)--\ref{fig:Susceptibility_QMC}(IIc)  
show that  as  $n_l$  grows  the  charge  susceptibility  remains activated in the 
low-temperature  limit.

\section{Discussion and Conclusion}\label{sec:conclusion}
The physics of  the half-filled  Hubbard and  Heisenberg models on  bipartite
lattices  has been investigated  in great  details  for Bravais lattices.   
Owing to  a theorem by Lieb \cite{Lieb89},  the  ground  state  on a  finite lattice 
is a  spin singlet.  The  lowest energy  state  in  each  total spin sector  builds  the Anderson tower of  states \cite{Anderson52},  which collapses in the 
thermodynamic limit.  This is  the mechanism  that  allows  for  the   broken-symmetry  quantum AFM   ground state  without violating  the   aforementioned  Lieb  theorem   in  dimensions   greater or  equal  to  two.  
In  one  dimension,   the  Heisenberg  and Hubbard  models,  are  critical  and 
are   described  by  an  SO(4) Wess-Zumino-Witten  theory.  

The notion of  thermodynamic limit  for  the hyperbolic lattices   differs  from
the  Euclidean case: for the  hyperbolic lattices  the  boundary dominates  over  the  bulk.   As  a  consequence,  the  very  notion of   dimensionality  and role 
of   fluctuations can be questioned.  While  our  mean-field  calculations  for  the Hubbard model  show  long-range  order on the boundary and  in the bulk,  a  spin-wave   calculation   for  the Heisenberg  model shows  that  fluctuations inhibit order  at  the  boundary.   As    demonstrated by our  QMC 
simulations  for  both the strong-coupling Hubbard and  Heisenberg models,  
this  turns  out  to be  an  artifact  of  the 
spin-wave  approximation.  It is  known  that spin waves  fail  to  predict that  weakly  coupling   chains  
with nonfrustrating  interactions is  a  relevant  perturbation \cite{Raczkowski13} that  leads to long range magnetic order.    
However, even if  the  ground state  is  ordered, it would  be of  interest  to 
investigate the  spin dynamical  response  on the  boundary, with the aim  to  assess 
whether  proximity  to one-dimensional  physics  is visible  at  high  frequencies
as  in Ref.~\cite{Raczkowski13}.   

Another   direction of  research is to  consider  frustration.  It  is known that weak frustrating  couplings  between  chains   do  not necessarily    lead to   magnetic ordering \cite{Starykh07},  thereby  offering  the possibility of realizing different   phases   in the bulk  and  on the boundary  of  hyperbolic 
lattices.

For the considered finite lattices with open boundary conditions, the $\{8,8\}$ lattice shows similarities to flat-band systems, with  a high total DOS at the Fermi level, which results in  the presence of AFM ordering at any finite interaction $U$.
The $\{8,3\}$ lattice displays odd-even effects with a vanishing (finite) DOS at the Fermi level for an even (odd) number of layers.
As for the honeycomb lattice, the Dirac-like character of the 
DOS close to the Fermi level of  the $\{10,3\}$ and $\{8,3\}$ lattices in the limit of large $n_l$ leads to a finite critical interaction strength $U_c$ for the onset of AFM ordering.
At  the  mean-field  level the  order parameter exponent takes  the  value $\beta = 1$  as  
for  the  Euclidean case.   On Euclidean  lattices,    great  progress   has  been achieved 
in the  understanding  of the Gross-Neveu  transition,   by  using renormalization 
group  invariant quantities  and finite-size scaling to  analyze  the data  \cite{Sorella12,Assaad13,Otsuka16}.  For the  hyperbolic   lattices  our 
analysis  is  very  rudimentary due  to the  very difficulty  of  defining a  continuum limit  and  the accompanying  critical  theory.    We  find this  to be  an interesting    line  of 
future research.  Nevertheless, our  data is consistent with a  finite-$U$  Mott  transition.

The observed  metal-insulator transition deserves more care since the bulk  and total DOS of the noninteracting system differ on hyperbolic lattices with open boundaries.  To define these quantities precisely,  let us consider a subset $M$ of bulk sites of the total number of sites, $N_s$.  The  bulk DOS is  given by
\begin{equation}
      N_{\text{Bulk}} (\omega)  =  - \frac{1}{\pi V_M}\sum_{i \in M }  \lim_{N_s \rightarrow \infty } 
      \text{Im} G^{R}_{i,i}(\omega)\,,
\end{equation}
where $G^{R}_{i,j}(\omega)$  is  the real-space  retarded Green function between site $i$ and $j$ and 
$V_M$  the number of sites in  the set $M$.   The  total DOS is defined  as in Eq.~(\ref{eq:dos}). We have shown explicitly that for the Dirac-like lattice, both bulk---with finite DOS at the Fermi energy---and   the total system with Dirac-like DOS do not show magnetic instabilities.    However, systems  with periodic boundary conditions, that have a finite DOS at the Fermi level and that are particle-hole symmetric, should show
magnetic instabilities and hence no metal-insulator transition.  In this sense,  the observed metal-to-insulator transition is \textit{boundary induced}.

The thermodynamic  properties  of the AFM Mott insulating state  are  the  very  same  
as  in the Euclidean case.  At low temperatures and in the   large  $n_l$  limit,    the 
charge  susceptibility is  activated  and the spin  susceptibility  constant.   We note  that 
the  notion of  Goldstone modes is  a property  of  the symmetry  group  and  not of  the
lattice  structure.  As  such it  is  not  surprising  to  observe   a finite spin  susceptibility in the low-temperature limit.

In the weak-coupling limit patterns of global AFM occur and can be observed in both mean-field and QMC results. Following the probability distribution of the low-lying energy eigenstates, macroscopic  ferromagnetic moments are formed, that are compensated in sum.  These patterns all have in common, that they break the $C_p$ 
(down to $C_{p/2}$) and time-reversal 
symmetry  of  the  Hamiltonian. Additionally the magnetic patterns are invariant under a combination of the  original rotation group of the lattice and time-reversal symmetry $C_p\mathcal{T}$. 
Global AFM occurs, e.g., also in finite graphene samples with zig-zag edges \cite{Fujita96,Fernandez07,Fernandez08,Bhowmick08,Feldner10,Yazyev10,Raczkowski17} or in more complex pattern, when the sample is subject to a strain \cite{Viana09,Roy14,Cheng15}.

The  absence  of  negative sign problem in QMC simulations hinges on 
symmetries (e.g.,  combined time-reversal and a  U(1)  conservation) \cite{Wu04,Li16}   that are   not  broken  by  the hyperbolic geometry.  Hence  all models  that are   amenable  to  negative sign-free  QMC simulations  such  as  SU(N) Hubbard-Heisenberg models \cite{Assaad04}, SU(N) Kondo lattices \cite{Assaad99a,Raczkowski20_a}, and   Su-Schrieffer-Heeger  Hamiltonians  \cite{Xing21,Cai21,Goetz22,Goetz23}  can be  studied.  It remains to be seen if novel  phenomena  can  be  observed   due  the 
hyperbolic  geometry.

\begin{acknowledgments}
  The authors thank Igor Boettcher and  Ronny  Thomale  for discussions.  
  Special  thanks  to  Igor Boettcher for providing a Mathematica notebook that generates adjacency matrices for hyperbolic lattices.
  The authors gratefully acknowledge the Gauss Centre for Supercomputing
  e.V. (www.gauss-centre.eu) for funding this project by providing computing
  time on the GCS Supercomputer SuperMUC-NG at the Leibniz Supercomputing Centre
  (www.lrz.de). F.A. and J.I. thank the W\"urzburg-Dresden
  Cluster of Excellence on Complexity and Topology in Quantum Matter ct.qmat
  (EXC 2147, project-id 390858490), G.R. and A.G. thank the DFG for financial support under the grant AS 120/19-1  (Project number 530989922).
  F.A.    acknowledges financial support from the German Research Foundation (DFG) under the grant AS 120/16-1 (Project number 493886309) that is part of the collaborative research project SFB Q-M\&S funded by the Austrian Science Fund (FWF) F 86.
\end{acknowledgments}

\appendix

\section{Spin-wave approximation}\label{sec:app_spin}
In this Appendix we briefly outline the calculation of the correction to the classical Néel state within the spin-wave approximation. Since the lattices do not possess translation symmetry, we choose an ansatz for the calculation along the lines of Refs.~\cite{Walker80,Sanyal12}. 
Inserting the ansatz of Eqs.~(\ref{eq:spin_a}) and $(\ref{eq:spin_b}$) into the Heisenberg Hamiltonian Eq.~(\ref{eq:heisenberg}) yields
\begin{equation}\label{eq:Heisen}
\hat{H} =S \sum_{\bm{i}} z_{\bm{i}} \hat{b}_{\bm{i}}^{\dagger}\hat{b}_{\bm{i}}^{\phantom{\dagger}} +\frac{S}{2} \sum_{\bm{i},\bm{j}} T_{\bm{i},\bm{j}}(\hat{b}_{\bm{i}}^{\phantom{\dagger}} \hat{b}_{\bm{j}}^{\phantom{\dagger}} +\hat{b}_{\bm{i}}^{\dagger} \hat{b}_{\bm{j}}^{\dagger}) - S^2 N_b \, .
\end{equation}
Here we defined the site-dependent coordination number $z_{\i} = \sum_{\j} T_{\i,\j}$. We can write the equations of motions for the boson operators in a matrix representation as 
\begin{equation}
\frac{d}{dt} \hat{\bm{b}} = i  
 \begin{pmatrix}
-P & -Q  \\
\bar{Q} & \bar{P}   \\
\end{pmatrix}
\hat{\bm{b}} \,,
\end{equation}
with
\begin{eqnarray}
 \hat{\bm{b}}^{\dagger} &=& ( \hat{b}_{\bm{i}=1}^{\dagger}, \dots,  \hat{b}_{\bm{i}=N_s}^{\dagger},\hat{b}_{\bm{i}=1}^{\phantom{\dagger}}, \dots,  \hat{b}_{\bm{i}=N_s}^{\phantom{\dagger}})\,, \nonumber \\
 P_{\i,\j} &=& S z_{\i} \delta_{\i,\j} \,, \quad Q_{\i,\j} = S T_{\i,\j} \,.
\end{eqnarray}
To diagonalize the Hamiltonian of Eq.~(\ref{eq:Heisen}) we perform a Bogoliubov transformation, 
\begin{equation}
 \begin{pmatrix}
-\Omega & 0  \\
0 & \Omega   \\
\end{pmatrix}=
 \begin{pmatrix}
\bar{g} &f  \\
\bar{f} &g   \\
\end{pmatrix}^{-1}
 \begin{pmatrix}
-P & -Q  \\
\bar{Q} & \bar{P}   \\
\end{pmatrix}
 \begin{pmatrix}
\bar{g} &f  \\
\bar{f} &g   \\
\end{pmatrix}\,.
\end{equation}
To ensure that the new boson operators $\hat{d}_{\i}^{\dagger}$, which we obtain from  $\hat{b}_{\i}^{\dagger} = \bar{f}_{\i,\j} \hat{d}_{\j}^{\phantom{\dagger}} + g_{\i,\j} \hat{d}_{\j}^{\dagger}$, still fulfill the canonical commutation relations, we need to impose a condition on the entries of $f$ and $g$:
\begin{equation}
 \begin{pmatrix}
\bar{g} &f  \\
\bar{f} &g   \\
\end{pmatrix}
 \begin{pmatrix}
-1& 0 \\
0 &1   \\
\end{pmatrix}
 \begin{pmatrix}
\bar{g} &f  \\
\bar{f} &g   \\
\end{pmatrix}^{\dagger}=
 \begin{pmatrix}
-1& 0 \\
0 &1   \\
\end{pmatrix} \,.
\label{eq:Bog}
\end{equation}
We now define a  scalar product  $\langle \bm{x},\bm{y} \rangle_B = \bm{x}^{\dagger}  \left(\begin{smallmatrix}
-1& 0 \\
0 &1   \\
\end{smallmatrix}\right) \bm{y}$.  The matrix $M=\left(\begin{smallmatrix}
-P&- Q \\
\bar{Q} & \bar{P}   \\
\end{smallmatrix}\right)$ is hermitian with respect to this scalar product  $\langle M \bm{x},\bm{y} \rangle_B=\langle  \bm{x},M\bm{y} \rangle_B$ 
and the Bogoliubov transformation $U= \left(\begin{smallmatrix}
\bar{g} &f  \\
\bar{f} &g   \\
\end{smallmatrix}\right)$ is unitary  $\langle U \bm{x},U\bm{y} \rangle_B=\langle  \bm{x},\bm{y} \rangle_B$. After diagonalizing the matrix $M$, the condition on the Bogoliubov transformation of Eq.~(\ref{eq:Bog}) can be satisfied by ensuring, that all eigenvectors of $M$ are orthonormal with respect to the scalar product $\langle \cdot,\cdot \rangle_B$. If the norm of an eigenvector $\langle \bm{x}_n,\bm{x}_n \rangle_B > 0$ is nonzero, then it can also be shown that the corresponding eigenvalue is real. To avoid eigenvectors with a vanishing $B$ norm we add an infinitesimal staggered magnetic field $\hat{H}_{\epsilon}=-\epsilon \sum_{\i} (-1)^{\i}\hat{S}_{\i}^z$. After the calculation, limits have to be taken carefully, by first taking the thermodynamic limit and then the limit of $\epsilon \rightarrow0$. The site-dependent correction to the classical AFM state can be calculated by the mean boson occupation number at $T=0$:
\begin{equation}\label{eq:1/S}
\langle \hat{S}_{\i}^z \rangle = S - \langle \hat{b}_{\i}^{\dagger}\hat{b}_{\i}^{\phantom{\dagger}} \rangle = S - \sum_{\j} |f_{\i,\j}|^2 \,.
\end{equation}

\section{Staggered spin susceptibility}\label{sec:suscep}

In this Appendix we  show that the  assumption of partial particle-hole symmetry and a finite DOS at the Fermi surface,  suffices to  argue a logarithmic divergence of the  staggered spin 
susceptibility.  This calculation does not depend on the lattice symmetry  and would  also hold for  disordered  systems  where the 
disorder does not break the said  symmetry.
The tight-binding part of the Hamiltonian is invariant under a partial particle-hole transformation defined as
\begin{equation}
\hat{P}^{-1}_{\sigma} \hat{c}_{\i,\sigma'}^{\dagger} \hat{P}_\sigma =
  \delta_{\sigma,\sigma'} (-1)^{\bm{i}}
  \hat{\gamma}_{\i,\sigma'}^{\phantom{\dagger}} + \left(1-\delta_{\sigma,\sigma'}
  \right) \hat{\gamma}_{\i,\sigma'}^{\dagger}\,.
\end{equation}
Without loss of generality we can consider the staggered spin susceptibility in the $x$ direction:
\begin{eqnarray}
\chi^{x}(Q) &=& \frac{1}{N_s}  \int_0^{\beta} d \tau \sum_{\bm{i},\bm{j}} (-1)^{\bm{i}+\bm{j}} \langle \hat{S}_{\bm{i}}^x (\tau) \hat{S}_{\bm{j}}^x \rangle \\
&=& \frac{2}{N_s}  \int_0^{\beta} d \tau \sum_{\bm{i},\bm{j}} \langle \hat{\gamma}_{\bm{i},\uparrow}^{\dagger} (\tau)\hat{\gamma}^{\phantom{\dagger}}_{\bm{j},\uparrow} \rangle   \langle \hat{\gamma}^{\dagger} _{\bm{i},\downarrow}(\tau) \hat{\gamma}^{\phantom{\dagger}}_{\bm{j},\downarrow} \rangle \, . \nonumber
\end{eqnarray}
As apparent,  the  partial  particle-hole transformation maps the   staggered   susceptibility  in the particle-hole  channel  to  the 
uniform  one in the  particle-particle channel. The  latter is nothing but  the  Cooper instability,  which we  now  show to be  present  on any graph with finite  DOS  at the Fermi energy. 
To do so,  we  introduce  a canonical transformation $U$, that diagonalizes the Hamiltonian 
\begin{equation}
\hat{H}_t =   \sum_{\bm{i},\bm{j},\sigma} h_{\bm{i},\bm{j}}  \hat{\gamma}_{\i,\sigma}^{\dagger}   \hat{\gamma}_{\j,\sigma}^{\phantom{\dagger}} = \sum_{x} \lambda_{x} \hat{\eta}_{x,\sigma}^{\dagger}   \hat{\eta}_{x,\sigma}^{\phantom{\dagger}} \,,
\end{equation}
with new fermionic operators $\hat{\eta}_{x,\sigma}^{\dagger}= \sum_{\i}  \hat{\gamma}_{\i,\sigma}^{\dagger} U_{\i,x}$. Applying this transformation yields 
\begin{eqnarray}
\chi^x(Q) = && \frac{2}{N_s} \int_0^\beta d\tau  \\
&& \times \sum_{x,y} f_{x,y} e^{\tau(\lambda_x+\lambda_y)}
\langle  \hat{\eta}_{x,\uparrow}^{\dagger}  \hat{\eta}_{x,\uparrow}^{\phantom{\dagger} } \rangle \langle  \hat{\eta}_{y,\downarrow}^{\dagger}  \hat{\eta}_{y,\downarrow}^{\phantom{\dagger} } \rangle
 \nonumber \\
= && \frac{1}{N_s} \sum_{x,y} \frac{f_{x,y}}{\lambda_x+\lambda_y} \left[\tanh{\left(\frac{\beta}{2}\lambda_x\right)} + \tanh{\left(\frac{\beta}{2}\lambda_y\right)} \right]\,, \nonumber
\end{eqnarray}
with $f_{x,y}=\sum_{\i,\j} (U^{\dagger})_{x,\i} U_{\j,x} (U^{\dagger})_{y,\i} U_{\j,y}$. For our specific case,  the hopping Hamiltonian is symmetric such that $U$  is an orthogonal matrix. As a consequence, $f_{x,y}= \delta_{x,y}$. Hence,
\begin{eqnarray}\label{eq:chi_x}
\chi^x(Q) & = &   \frac{1}{N_s} \sum_{x} \frac{ 1}{\lambda_x} \tanh{\left(\frac{\beta}{2}\lambda_x\right)}    \\
&=& \int d \epsilon \rho(\epsilon) \frac{1}{\epsilon} \tanh{\left(\frac{\beta}{2}\epsilon\right)}   \approx  \rho(0) \ln \left(\frac{W}{2 k_b T}\right)  \, , \nonumber
\end{eqnarray}
where  the DOS is given by Eq.~(\ref{eq:dos}).
The last approximation is in the low-temperature limit and $W$ is a high-energy cutoff. Provided that the DOS is finite at the Fermi energy, an instability toward AFM ordering is  generic even in the absence of translational symmetry.

\section{Instability of the bulk}\label{sec:bulk}
To study the stability of the bulk toward AFM ordering, we place an AFM perturbation $\hat{H}_I$ on the inner ring of the lattice
\begin{eqnarray}\label{eq:bulk_instability}
\hat{H} &=& \hat{H}_t + \hat{H}_I \,, \\
\hat{H}_t &=&   \sum_{i,i', n,n', \sigma}  \hat{c}^{\dagger}_{i,n,\sigma} T_{n,n'}(i-i')\hat{c}^{ \phantom\dagger}_{i',n',\sigma} \,, \nonumber \\
\hat{H}_I &=&   \sum_{i,\sigma} m \sigma (-1)^{i}   \hat{c}^{\dagger}_{i,1,\sigma}  \hat{c}^{\phantom\dagger}_{i,1,\sigma} = \sum_{k,\sigma}m \sigma  \hat{c}^{\dagger}_{k,1,\sigma}  \hat{c}^{\phantom\dagger}_{k+Q,1,\sigma} \,.\nonumber
\end{eqnarray}
In Eq.~(\ref{eq:bulk_instability}), we used the  notation as introduced in Eq.~(\ref{eq:altermagnet}), where we defined a  unit cell $i$ with the help of the $C_p$ symmetry of the lattice. The orbital $n=1$ refers to the site on the inner layer included in the unit cell $i$. Equation~(\ref{eq:bulk_instability}) is equivalent to Eq.~(\ref{eq:altermagnet}) by setting $M_{1,1}=m$ and $M_{n,n'}=0$ otherwise.

The susceptibility with respect to the perturbation $\hat{H}_I$ can be written as
\begin{eqnarray}
\chi(Q,i\Omega_m) &=& \int_0^{\beta} d\tau e^{i \Omega_m \tau} \langle \hat{H}_I(\tau) \hat{H}_I\rangle_0 \nonumber \\
&=& - 2 m^2 \int_0^{\beta} d\tau e^{i \Omega_m \tau} G_{1,1}(k,-\tau) G_{1,1}(k+Q,\tau) \nonumber \\
&=& 2 m^2 \sum_k \int d\omega d\omega'   \frac{f(\omega')-f(\omega)}{\omega-\omega'+i\Omega_m} \nonumber\\
&&\times A_{1,1}(k,\omega)A_{1,1}(k+Q,\omega') \,.
\end{eqnarray}
Here, we defined the Green function $G_{1,1}(k,\tau)=-\langle\mathcal{T}\hat{c}_{k,1}^{\phantom{\dagger}}(\tau)\hat{c}_{k,1}^{\dagger}\rangle_0$ with the imaginary-time ordering operator $\mathcal{T}$ and the expectation value $\langle\dots\rangle_0$ with respect to $\hat{H}_t$. The spectral function $A_{1,1}(k,\omega)=-\frac{1}{\pi}\mathrm{Im}G_{1,1}(k,i\omega_m\rightarrow \omega+i\delta)$ is defined via analytical continuation of the Green function with an infinitesimal $\delta$ and the fermionic Matsubara frequencies $\omega_m$. Using particle-hole symmetry, $A_{1,1}(k,\omega) =  A_{1,1}(k + Q,-\omega) $,  we  can  further simplify  the last equation  to  obtain:
\begin{eqnarray}
  \chi(Q,i\Omega_m) & =  & 2 m^2  \int d\omega d\omega'   \frac{f(-\omega')-f(\omega)}{\omega+\omega'+i\Omega_m}  \nonumber \\
   & &\times \sum_k A_{1,1}(k,\omega)A_{1,1}(k,\omega') \,.
\end{eqnarray}
Since $A_{1,1}(k,\omega)$ takes a finite, constant value  at the  Fermi  energy \cite{Mosseri23}  we  can 
replace  the  $k$-sum over  the spectral functions  by a finite number.   Carrying out the  integration 
leads to a finite value  of  $ \chi(Q,i\Omega_m = 0) $ such that no bulk magnetic  instability is present.

\section{Additional data}\label{sec:additional}
In the following Appendix we present some additional data.
\subsection{Mean-field approximation}\label{sec:add_mean}
 \begin{figure}[tb]
 \includegraphics[width=1\linewidth]{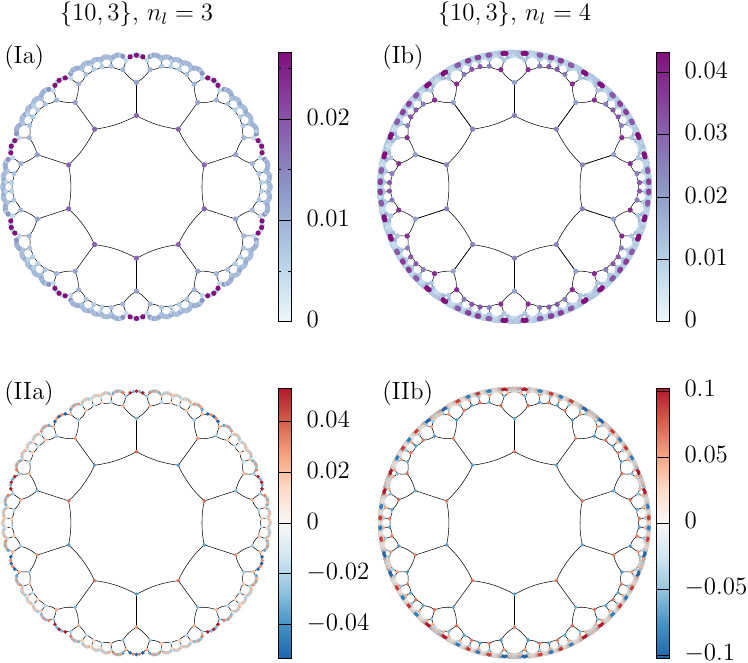}
  \caption{\label{fig:global_afm_p10_q3}(I) Local DOS $\rho(E=0,\bm{i})$ at zero energy in the noninteracting limit $U=0$ at $\delta=0.01$ and (II) magnetization $m_{\bm{i}}^z$ as obtained from mean-field approximation for the $\{10,3\}$ lattice and interaction strength $U=0.8t$. }
\end{figure}
 \begin{figure}[tb]
 \includegraphics[width=1\linewidth]{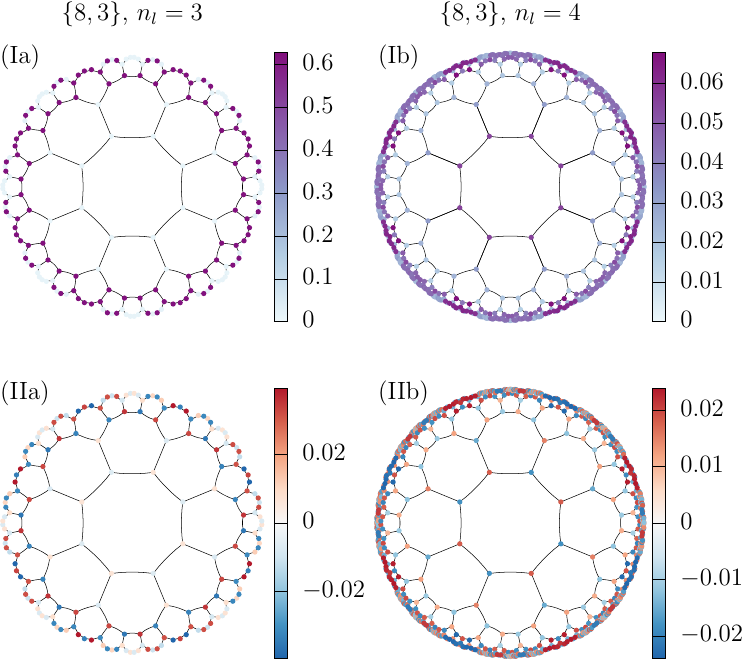}
  \caption{\label{fig:global_afm_p8_q3}Same as in Fig.~\ref{fig:global_afm_p10_q3}, but for the $\{8,3\}$ lattice and interaction strength (IIa) $U=0.4t$ and (IIb) $U=0.58t$. }
\end{figure}
\begin{figure}[htb]
 \includegraphics[width=1\linewidth]{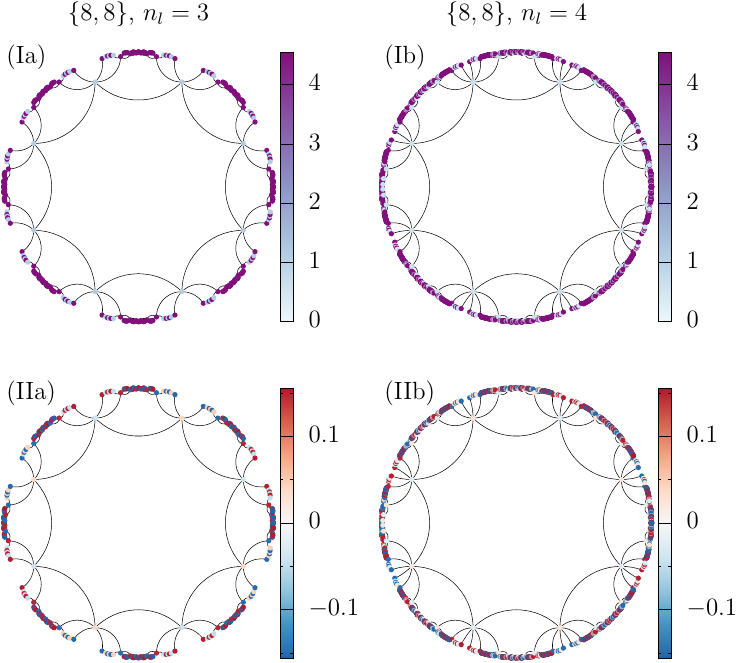}
  \caption{\label{fig:global_afm_p8_q8}Same as in Fig.~\ref{fig:global_afm_p10_q3}, but for the $\{8,8\}$ lattice and interaction strength $U=0.1t$.  }
\end{figure}

 In Sec.~\ref{sec:mean_field} we showed the spatial distribution of the magnetization $m_{\bm{i}}^z$ in the weak-coupling limit for the $\{8,3\}$ lattice with $n_l=5$ layers. In the upper panels of Figs.~\ref{fig:global_afm_p10_q3}--\ref{fig:global_afm_p8_q8} we plot the local DOS at zero energy, 
\begin{equation}
\rho(E,\bm{i}) = -\frac{1}{\pi N_s}  \sum_n |U_{\bm{i},n}|^2  \mathrm{Im} G^{R}(n,E)\,, 
\end{equation}
for the different lattice geometries both with  $n_l=3$ and $n_l=4$ layers.
Here we defined a canonical transformation $U$, that diagonalizes the mean-field Hamiltonian~(\ref{eq:Ham_MF}).
 They all show a unique pattern of low and high DOS, which is compatible with the $C_{p}$ symmetry of the lattices.
 The lower panels of Figs.~\ref{fig:global_afm_p10_q3}--\ref{fig:global_afm_p8_q8} show the corresponding local order parameter in real space, where again local ferromagnetic and AFM moments emerge in the  different sectors of the lattice. But in  sum of all lattice sites only the AFM moment remains.

 \subsection{Spin-wave approximation}\label{sec:add_spin}
\begin{figure*}[htb]
 \includegraphics[width=\linewidth]{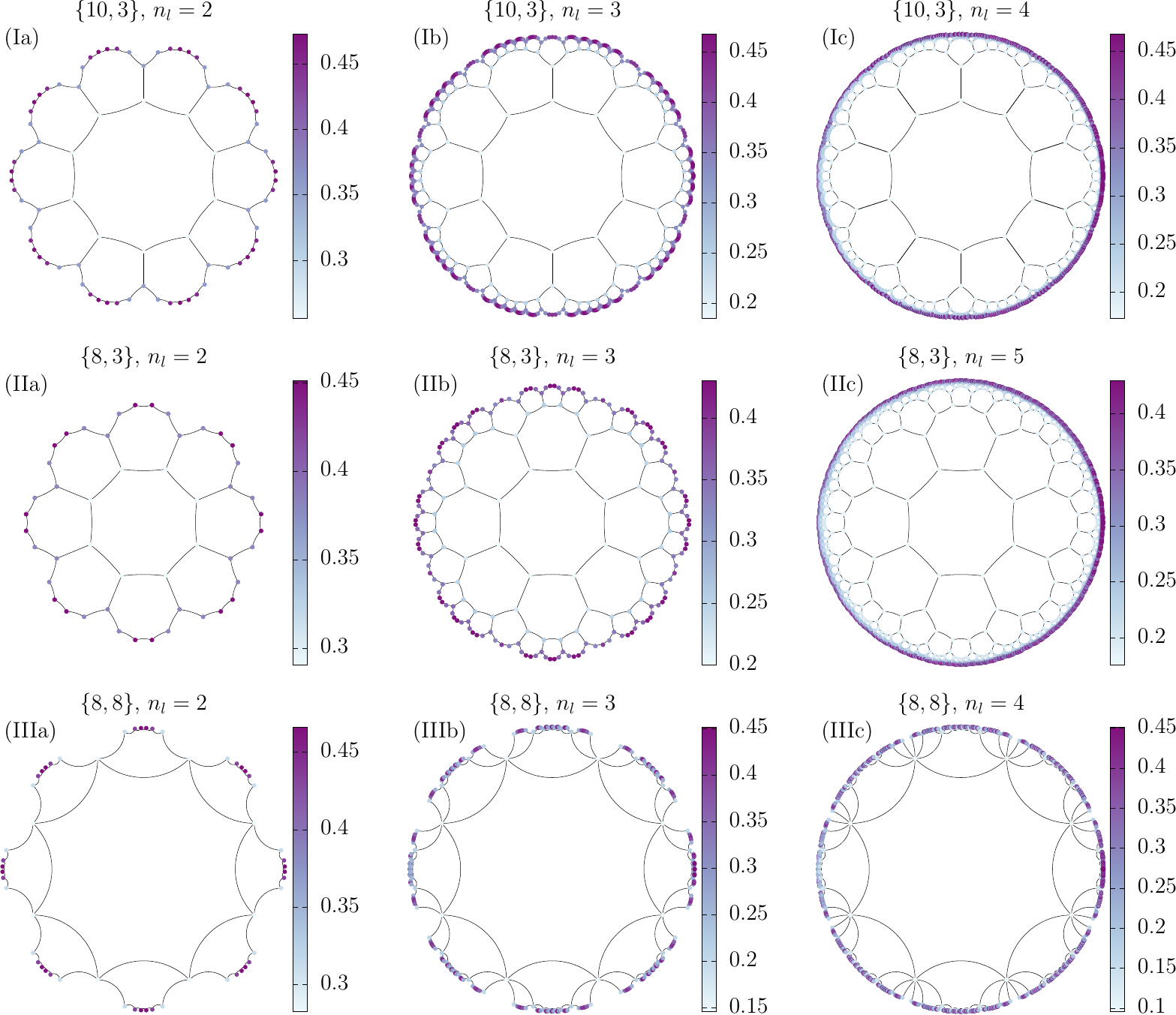}
  \caption{\label{fig:spin_wave_more_latt}Local correction $\langle \hat{b}_{\bm{i}}^{\dagger}\hat{b}_{\bm{i}}^{\phantom{\dagger}}\rangle$ to classical Néel state as obtained from spin-wave approximation for the (I) $\{10,3\}$, (II) $\{8,3\}$, and (III) $\{8,8\}$ lattice and varying system size $n_l$ at $\epsilon=0.01$.}
\end{figure*}

 The results for the site-dependent correction $\langle \hat{b}_{\i}^{\dagger} \hat{b}_{\i}^{\phantom{\dagger}}\rangle$ to the classical Néel state obtained from the spin-wave approximation for the $\{10,3\}$ and $\{8,3\}$ lattice in Figs.~\ref{fig:spin_wave_more_latt}(I) and \ref{fig:spin_wave_more_latt}(II) are similar to the results for the $\{8,3\}$ lattice described in the main text [Fig.~\ref{fig:spin_wave_local}]. For the $\{8,8\}$ lattice the results differ slightly due to the lattice geometry. The correction  on many sites in the outermost layer is smaller than for the other two lattices, since  the distance from one site on the edge to another site on the edge is in general smaller. This is a consequence of the high coordination number $q=8$,  since  sites in the bulk can have direct connections to the edge. The sites on the edge also form short one-dimensional chains, but larger effective interactions are present in comparison to the other two considered lattice geometries.
 
\subsection{Quantum Monte Carlo}
\begin{figure*}[htb]
	\includegraphics[width=\linewidth]{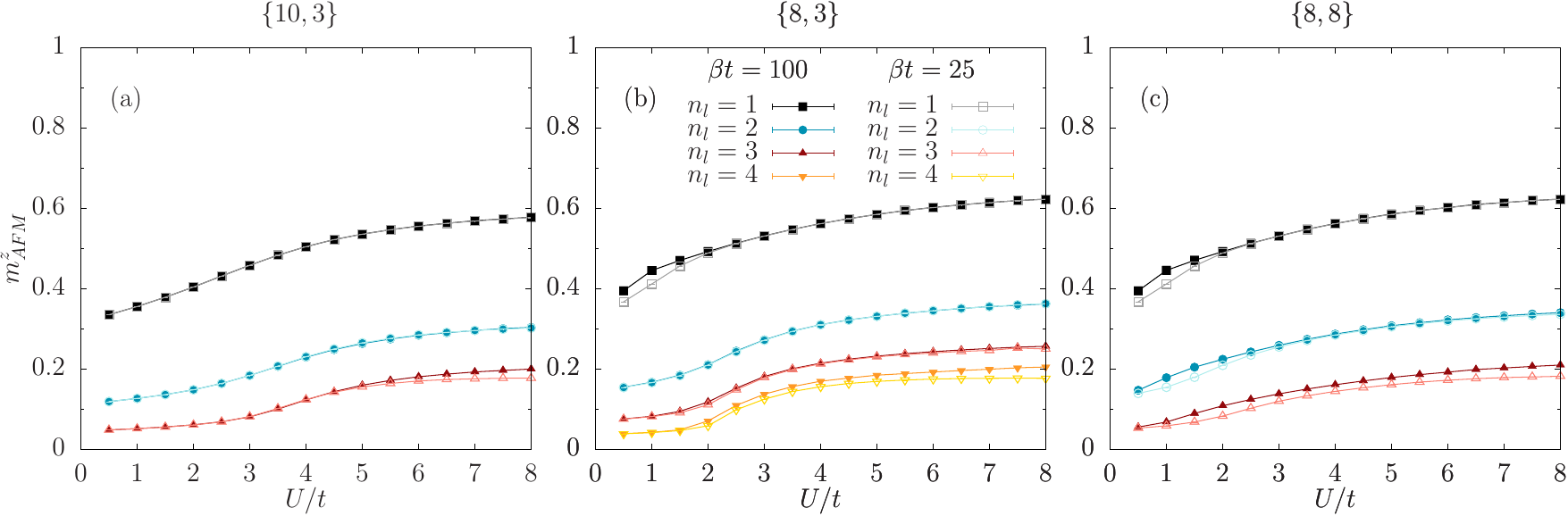}
	\caption{\label{fig:m_U} AFM order parameter for the $\{10,3\}$, $\{8,3\}$, and $\{8,8\}$ lattice obtained by QMC calculations at $\beta t=100$ (filled symbols) and $\beta t=25$ (unfilled symbols).}
\end{figure*}

To supplement the picture given in Figs.~\ref{fig:Mag_afm_QMC} and  \ref{fig:QMC_global_AF},  we present the order parameter $m_{\mathrm{AFM}}^z$ as a function of $U$ in Fig. \ref{fig:m_U} for two different temperatures $\beta t=25$ and $\beta t=100$. We  see that the  $\{8,8\}$ lattice exhibits stronger temperature effects than the other lattices. This lattice shows a strong sensitivity to temperature already in the weak coupling regime, as  can be seen in Fig. \ref{fig:m_U}(c). As discussed before, this effect is also present in the  results from a mean-field analysis  [Fig. \ref{fig:mag_temp}(a)].

\clearpage
\bibliography{hyper}

\end{document}